\documentclass{elsart}
\usepackage{amsmath,amssymb,graphicx}

\def\){\right)}
\def\({\left(}
\newcommand{\be}{\begin{equation}}
\newcommand{\ee}{\end{equation}}
\newcommand{\la}{\Lambda}
\newcommand{\vlk}{V_{\text{low k}}}
\newcommand{\vlkb}{\overline{V}_{\text{low k}}}

\begin{document}
\begin{frontmatter}

\title{Model-independent low momentum nucleon interaction
from phase shift equivalence}
\author{S.K. Bogner$\hspace{1mm}^{a,b}$\thanksref{SKB}},
\author{T.T.S. Kuo$\hspace{1mm}^{b}$\thanksref{TTSK}} and
\author{A. Schwenk$\hspace{1mm}^{c,b}$\thanksref{AS}}
\address{$^{a}$Institute for Nuclear Theory, Box 351550, University
of Washington,\\
Seattle, WA 98195\\
$^{b}$Department of Physics and Astronomy, State University of
New York,\\
Stony Brook, NY 11794-3800\\
$^{c}$Department of Physics, The Ohio State University, Columbus, 
OH 43210}

\thanks[SKB]{E-mail: bogner@phys.washington.edu}
\thanks[TTSK]{E-mail: kuo@nuclear.physics.sunysb.edu}
\thanks[AS]{E-mail: aschwenk@mps.ohio-state.edu}

\begin{abstract}

\noindent
We present detailed results for the model-independent low momentum
nucleon-nucleon interaction $V_{\text{low k}}$. By introducing a
cutoff in momentum space, we separate the Hilbert space into a low
momentum and a high momentum part. The renormalization group is used 
to construct the effective interaction $V_{\text{low k}}$ in the low 
momentum space, starting from various high precision potential models 
commonly used in 
nuclear many-body calculations. With a cutoff in the range of $\la
\sim 2.1 \, \text{fm}^{-1}$, the new potential $V_{\text{low k}}$ is 
independent of the input model, and reproduces the 
experimental phase shift data for corresponding laboratory energies 
below $E_{\text{lab}} \sim 350 \, \text{MeV}$, as well as the deuteron 
binding energy with similar accuracy as the realistic input 
potentials. The model independence of $V_{\text{low k}}$ demonstrates
that the physics of nucleons interacting at low momenta does not 
depend on details of the high momentum dynamics assumed 
in conventional potential models. $V_{\text{low k}}$ does not have 
momentum components larger than the cutoff, 
and as a consequence is considerably softer than the high 
precision potentials. Therefore, when $V_{\text{low k}}$ is used 
as microscopic input in the many-body problem, the high momentum 
effects in the particle-particle channel do not have 
to be addressed by performing a Brueckner ladder resummation or 
short-range correlation methods. By varying the cutoff, we study how 
the model independence of $V_{\text{low k}}$ is reached in different 
partial waves. This provides numerical evidence for the separation of 
scales in the nuclear problem, and physical insight into the nature of
the low momentum interaction.

\vspace{0.5cm}

\noindent{\it PACS:}
13.75.Cs;	
21.30.-x;	
11.10.Hi \\     
\noindent{\it Keywords:} Nucleon-Nucleon Interactions; Effective
Interactions; Renormalization Group
\end{abstract}
\end{frontmatter}

\section{Introduction}

A major challenge of nuclear physics lies in the description
of finite nuclei and nuclear matter on the basis of a microscopic
theory. By microscopic, it is understood that the input
nuclear force is, as best as possible, based on the available free-space
data, e.g., the elastic nucleon-nucleon and nucleon-deuteron
scattering phase shifts. The theoretical predictions are then obtained using
a systematic many-body approach, which is reliable for strongly interacting 
systems.

At low energies and densities, one typically starts by treating 
protons and neutrons as nonrelativistic point-like fermions. However, as 
opposed to the Coulomb interaction in electronic systems, the
nucleon-nucleon interaction remains unknown from the underlying
theory of the strong interactions, Quantum Chromodynamics (QCD).
Thus, phenomenological meson-exchange models of the
two-nucleon force are commonly used as input in many-body calculations.
Moreover, three- (and possibly four-) body forces are needed 
to obtain accurate results for few-body systems. Recent Green's 
function Monte Carlo calculations using a realistic 
two-body interaction, the Argonne $v_{18}$ potential, have
shown that three-body forces typically contribute a net attraction
of $\approx 20 \%$ to the low-lying states
of light nuclei, $A \leqslant 10$~\cite{GFMC,GFMC9+10}.

However, already at the lowest level of the 
two-body interaction, there are several quite different 
potential models commonly used. Therefore, it is a
natural first step to ask whether it is possible to remove or to
minimize this model dependence. In this work, we expand on previous
results~\cite{Vlowk,Vlowkflow} and demonstrate that this is
indeed possible, as long as one restricts the interaction to
momenta which are constrained by the experimental scattering data,
and explicitly includes the long-range part of the interaction.

There are a number of high precision models of the nucleon-nucleon force
$V_{\text{NN}}$, which we will refer to as the ``bare'' interactions. 
All realistic potentials are given by the one-pion exchange (OPE) 
interaction at long distances, but vary substantially in their treatment of
the intermediate-range attraction and the short-range repulsion.
We briefly summarize the different frameworks used to generate
the intermediate and short-range parts of the realistic
potential models~\cite{NN1,NN2}. The Paris potential~\cite{Paris} 
calculates the intermediate-range contributions to the nucleon-nucleon 
scattering
amplitude from two-pion exchange using dispersion theory, and then
assumes a local potential representation consisting of several
static Yukawa functions. 
The $\omega$-meson exchange at short distances is included as part of the
three-pion exchange, and a repulsive core is introduced by sharply cutting
off these parts at the internucleon distance $r \sim 0.8\,\text{fm}$. At short
distances, the interaction is simply given by a constant (but 
energy-dependent) soft core. The Bonn potentials~\cite{Bonn} are based 
on multiple one-boson exchange interactions and the two-pion exchange 
potential calculated in perturbation theory. The two-pion exchange 
contribution is then approximated by an energy-independent $\sigma$-meson
exchange term. Smooth form factors (with typical 
form factor masses $\approx 1-2 \,
\text{GeV}$) cut off the potential at short distances. The short range
repulsion originates from the $\omega$ exchange. The second generation
CD-Bonn potential~\cite{CDBonn1,CDBonn2} refines the treatment of the
one-boson exchange amplitudes to take into account the
non-local structure arising from the covariant amplitudes. The Nijmegen 
potentials~\cite{Nijmegen}
consist of multiple one-boson exchange parts (with separate local and non-local versions), where the interaction parameters
of the heavier mesons depend on the partial waves. At very short 
distances, the potentials are regularized by exponential form factors.
Finally, the local Argonne potential~\cite{Argonne} consists of the OPE 
interaction regularized at short distances,
and a phenomenological parametrization at short and 
intermediate distances. The core is provided by Woods-Saxon functions 
that are effective at a distance $r \sim 0.5\,\text{fm}$.

Over the past decade, there has also been a great effort in deriving
low-energy nucleon-nucleon interactions in the framework of effective field
theory (EFT)~\cite{Weinberg1,Weinberg2,vanKolck,KSW,Park,EGM,Idaho} (for a
review see~\cite{encyc}). The EFT approach is based on local 
Lagrangian field theory with low-energy degrees of freedom,
constrained by the symmetries of QCD. The Lagrangian contains 
the nucleon and pion fields and all possible interactions
consistent with chiral symmetry. At low energies, the heavy mesons and
nucleon resonances can be integrated out of the theory. Their effects
are encoded in the renormalized pion exchange and scale-dependent
coupling constants of model-independent contact interactions.
By formulating  a power counting scheme, the number of
couplings in the effective Lagrangian can be truncated and fitted to a
set of low-energy data. Once the low-energy couplings are determined,
the power counting scheme enables the EFT to predict other processes
with controlled error estimates. 

In addition to the high precision 
phenomenological interactions, we study the Idaho
potential~\cite{Idaho}, which is a chiral model inspired by EFT.
The Idaho potential introduces some model dependence in contrast to
the rigorous EFT interactions by, e.g., including several terms in the 
expansion which are subleading in the power counting with respect to 
omitted ones. This leads to a better description of the experimental
phase shifts, comparable in accuracy to the realistic models. 
Since we will argue that our results are dependent
on the accurate reproduction of the scattering phase shifts, the
results for the Idaho EFT potential are presented here as well.

\begin{figure}[t]
\vspace{-8mm}
\begin{center}
\includegraphics[scale=1.15,clip=]{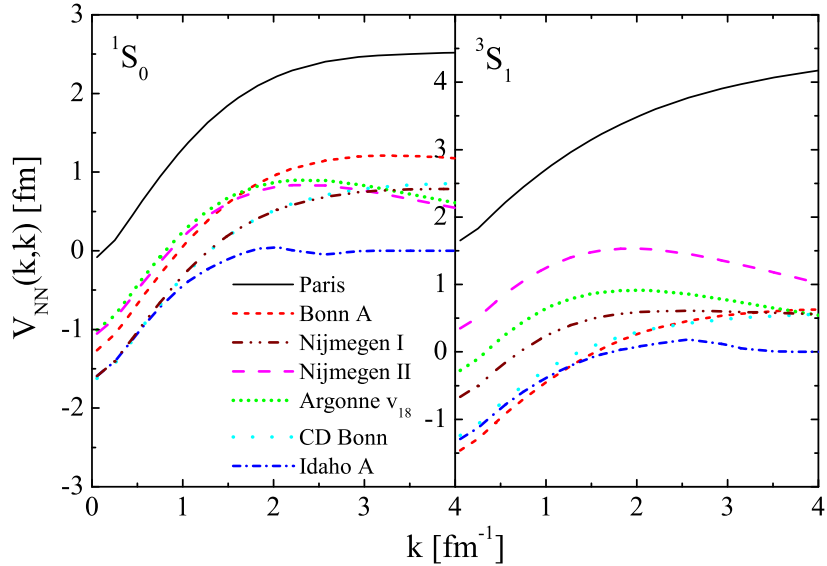}
\end{center}
\vspace{-8mm}
\caption{The diagonal momentum-space matrix elements of the different 
high precision potentials $V_{\text{NN}}$ versus relative momentum 
in the $^1$S$_0$ and $^3$S$_1$ partial wave.}
\label{barecomp}
\end{figure}
The couplings and parameters of the different potentials described
above are fitted to the elastic nucleon-nucleon scattering phase shifts
over laboratory energies $E_{\text{lab}} \lesssim 350 \, \text{MeV}$
and the low-energy deuteron properties. As a consequence, the 
bare interactions are constrained by the experimental data
up to a corresponding relative momentum scale of\footnote{Here,
we use conventional scattering units where $c = \hbar = \hbar^2/m =
1$, i.e., the laboratory energy is related to the relative momentum
through $\displaystyle E_{\text{lab}} = 2 \, k^2$.}
\be
\la_{\text{data}} \sim 2.1\,\text{fm}^{-1} .
\ee
This is manifest in Fig.~\ref{barecomp}, where we observe that various high
precision potential models of $V_{\text{NN}}$ have quite different 
momentum-space components, despite their common treatment of the OPE 
interaction and the reproduction of the same low-energy data. This 
indicates that the low-energy observables are not sensitive to the 
details or different assumptions of the short-distance physics. 
This insensitivity is a consequence of the separation of long and 
short-distance scales in the nuclear force, and implies that 
the nucleon-nucleon interaction is amenable to an effective theory or
renormalization group treatment.

\subsection{The two-body problem in effective theories}

In this work, we propose a renormalization group (RG) approach to the 
nucleon-nucleon interaction which removes the short-distance 
model dependencies of the high precision potentials, while preserving 
their high accuracy description of the nucleon-nucleon 
scattering data and deuteron properties. As our approach 
is a compromise between an EFT treatment and the high 
precision potentials, we first discuss in general terms how 
effective theory methods can be applied to the two-nucleon problem. 

The fundamental principle underlying modern effective theory 
methods is simple: the low-energy physics is not sensitive to the 
details of the high-energy dynamics. In other words, one can construct 
arbitrarily many theories, which have the same long-wavelength 
structure and lead to identical low-energy observables, but differ 
at short distances and higher-energy scales. This ambiguity can be
used constructively. Namely, it is possible to replace the detailed 
short-distance dynamics by simple effective interactions,
which preserve the low-energy symmetries of the full theory.

For example, a forerunner to the modern EFT approach is the well-known 
effective range theory of Bethe and Longmire~\cite{effrange}. This 
allowed a quantitative description of low-energy nuclear phenomena without 
a detailed knowledge of the nuclear force. The low-energy 
scattering amplitude $T(k,k;k^2)$ is given by two 
phenomenological low-energy constants independent of an assumed 
nuclear force model~\cite{NNInt}
\be
\frac{1}{T(k,k;k^2)} = \frac{1}{a_{s}} - \frac{1}{2} \, r_{0} \, k^2 +
\mathcal{O}(k^4) ,
\label{ERT}
\ee 
where $a_{s}$ denotes the scattering length and $r_{0}$ is the effective 
range parameter. The effective range theory encompasses the fundamental 
principle of modern effective theories, as any force models with 
different short-distance details can be tuned to give the same 
low-energy physics, which is captured in the effective range 
parameters.

More generally, this insensitivity can be exploited when the
short-distance dynamics are either poorly understood, or are too
complicated for calculation. Consider a quantum system
described by the Hamiltonian
\be 
H = H_{0} + V_{\text{L}} + V_{\text{H}} ,
\label{genericH}
\ee 
where $V_{\text{L}}$ primarily couples to low-energy
states, in our case the long-range part of the 
OPE interaction, and $V_{\text{H}}$ includes the remaining 
complicated or unknown short-distance part of the interaction.

A first step in building an effective theory is to impose a
cutoff $\Lambda$ on the intermediate state energies and momenta. In
the modern viewpoint, the cutoff takes on a physical meaning. It
divides the low-energy states, which are essential to the low-energy
physics, from the high-energy modes. In this way, one explicitly
maintains only dynamics that are well-understood. It is, however, 
not possible to simply neglect the remaining $V_{\text{H}}$, since 
its effects on the low-energy physics need to be included in the 
form of correction, or so-called counterterms. The correction terms 
are constrained by demanding that all low-energy spectra, low-energy 
amplitudes, etc., calculated using the bare Hamiltonian $H$ are 
exactly reproduced using an effective Hamiltonian $H_{\text{low k}}$ 
in the truncated Hilbert space. 

Schematically, in perturbation theory all amplitudes are of the form
\be 
\langle f | \mathcal{A} | i \rangle = \langle f | V_{\text{bare}} | i
\rangle + \sum_{n=0}^{\infty} \frac{\langle f | V_{\text{bare}} |
n \rangle \langle n | V_{\text{bare}} | i \rangle}{E_{i}-E_{n}} +
\mathcal{O}(V_{\text{bare}}^3) ,
\ee 
where the intermediate state summations are over both low and
high-energy states. We now define an effective low momentum potential 
$V_{\text{low k}}$ by demanding
\be
\langle f | \mathcal{A} | i \rangle = \langle f | V_{\text{low k}} | i
\rangle + \sum_{n=0}^{\Lambda} \frac{\langle f | V_{\text{low k}} |
n \rangle \langle n | V_{\text{low k}} | i \rangle}{E_{i}-E_{n}} +
\mathcal{O}(V_{\text{low k}}^3) ,
\ee
where we introduce the correction terms through
\be
V_{\text{low k}} = V_{\text{L}} + \delta V_{\text{ct}} .
\ee
The physical amplitudes cannot depend on the way in which we 
split the Hilbert space and must therefore be independent of 
the cutoff. Consequently, $V_{\text{low k}}$ changes or ``runs'' with the
cutoff in order to cancel the cutoff dependence 
arising from the truncated intermediate state summations. This means
that the effective theory must be RG invariant. Demanding RG
invariance implies a RG equation for the low momentum interaction
\be 
\frac{d}{d \Lambda} \langle f | \mathcal{A} | i \rangle = 0
\quad \Longrightarrow \quad \frac{d}{d \Lambda} V_{\text{low k}} =
\beta \bigl([V_{\text{low k}}],\Lambda\bigr) ,
\label{generalRGE}
\ee
where the $\beta$ function of the problem depends on the
cutoff-dependent low momentum potential, as well as explicit cutoff 
dependence from the regularization. Starting from a bare
interaction defined in the full Hilbert space, i.e., with a large
cutoff, the RG equation can be solved to obtain the physically equivalent 
effective potential. This process is referred to as integrating out, or 
decimating the high-energy degrees of freedom. It can be viewed as 
filtering out the details of the assumed high-energy dynamics, while 
implicitly incorporating their detail-independent effects on the 
low-energy physics.

Since the virtual high-energy states propagate over distances of the 
order $1/\Lambda$ or shorter, the correction terms are well approximated
by contact interactions, regardless of the detailed form of 
$V_{\text{H}}$. The general form of $\delta V_{\text{ct}}$ should 
therefore be given by a sum over all local contact operators
consistent with the symmetries of the problem. In the specific case 
of the two-body system, we have schematically 
(for an excellent discussion of the details of the approach, 
see Lepage~\cite{Lepage}.)
\begin{eqnarray}
\delta V_{\text{ct}}({\bf r}) &=& C_{0} \, \frac{1}{\Lambda^2} \,
\delta^3({\bf r}) + C_{2} \, \frac{1}{\Lambda^4} \, \nabla^2 \, 
\delta^3({\bf r}) + C_{2}' \, \frac{1}{\Lambda^4} \, {\bf\nabla} \cdot
\delta^3({\bf r}) \, {\bf\nabla} + \ldots \nonumber \\ 
&&+ \: C_{2n} \, \frac{1}{\Lambda^{2n+2}} \, \nabla^{2n} \,
\delta^3({\bf r}) + \ldots ,
\label{contactterms}
\end{eqnarray} 
with dimensionless couplings $C_{2n}$. To a low-energy probe, the 
effective theory is indistinguishable from the underlying theory, 
provided the decimation has been performed exactly.

The power counting scheme of the EFT enables one to reliably 
truncate the correction terms, Eq.~(\ref{contactterms}), and fit the
couplings directly to a set of low-energy data, since a RG decimation 
of the underlying fundamental theory is at present not feasible. 
In our current approach, we will
assume that the nucleon-nucleon interaction in the full Hilbert space
is given by a high precision potential model, and then perform the RG
decimation to low momenta exactly. In this way, the resulting 
low momentum potential contains all necessary counterterms to maintain 
exact RG invariance of the theory, and therefore reproduces the
experimental phase shift data and the deuteron properties over
the same kinematic range as the conventional models.

\subsection{A schematic model}

Before considering realistic nucleon-nucleon potential models, we apply 
the RG approach to a simplified schematic model that allows
for a straightforward solution of the RG equation to construct 
$V_{\text{low k}}$. In addition, the schematic model nicely
illustrates the main point of our current approach. Independent 
of the schematic model constructed in the full Hilbert space,
the RG decimation leads to a model-independent low momentum 
interaction $V_{\text{low k}}$.

We consider a separable interaction, which we construct to approximately 
fit the experimental neutron-proton phase shift data in
the $^1$S$_0$ channel. For this purpose, we found the following
separable interaction to be particularly useful\footnote{For
simplicity, we have chosen different functions for $L(k')$ and $R(k)$,
in order to include a repulsive part in the potential without having
to introduce several separable terms, which would complicate the
discussion.}
\begin{align}
V_{\text{bare}}(k',k) & = L(k') \: R(k) \\
& = \frac{g}{(k^{\prime 2}+m^2)^n} \biggl( - \frac{g}{(k^2+m^2)^n} +
\frac{\alpha \, g \, (k^2+m^2)^n}{(k^2+\eta^2 \, m^2)^{2n}} \biggr) .
\label{schem}
\end{align}
For the diagonal matrix elements, the analogy to the nucleon-nucleon
interaction can be seen better
\be
V_{\text{bare}}(k,k) = - \frac{g^2}{(k^2+m^2)^{2n}} + \frac{\alpha \,
g^2}{(k^2+\eta^2 \, m^2)^{2n}} .
\ee
The separable model consists of an attractive part and a short-range,
$\eta > 1$, repulsive part. The strength of the repulsion, for given
$\eta$, can be constructed to fit the $^1$S$_0$ phase shift
data reasonably.

For separable potentials, the Lippmann-Schwinger equation can be
resummed explicitly and one finds~\cite{NNInt}
\be
T_{\text{bare}}(k',k;k^2) = \cfrac{L(k') \, R(k)}{1 + \cfrac{2}{\pi}
\, \mathcal{P} \int\limits_{0}^{\infty} \cfrac{L(p) \,
R(p)}{p^{2}-k^{2}} \, p^{2} dp} .
\ee
We choose a typical $\eta = 2$, which approximately reflects the ratio of the
repulsive to attractive meson contributions $m_{\omega}/ 2 m_\pi$.
The remaining parameters of our model, $g, m$ and $\alpha$,
are then fitted to reproduce the effective range expansion,
\be
T_{\text{bare}}(k,k;k^2) = \frac{1}{\frac{1}{a_{s}} - \frac{1}{2} \, r_{0}
\, k^2} ,
\ee
and the change in sign of the experimental phase shifts $T_{\text{bare}}
(\kappa,\kappa;\kappa^2)=0$ at $\kappa = 1.79 \, \text{fm}^{-1}$
for a given power $n$. We constructed two exemplary separable models 
with the parameters given in Table~\ref{schempar}.
\begin{table}[t]
\begin{center}
\begin{tabular}{@{\hspace{0.4cm}}c@{\hspace{0.8cm}}c@{\hspace{0.8cm}}c@{\hspace{0.8cm}}c@{\hspace{0.4cm}}}
\hline
& I & II \\
\hline\hline
$n$ & 1/2 & 1 \\
$\alpha$ & 1.8199 & 4.5354 \\
$g \: [\text{fm}^{(1-4n)/2}]$ & 3.1429 & 3.3542 \\
$m \: [\text{fm}^{-1}]$ & 1.0962 & 1.3892 \\
\hline
\end{tabular}
\vspace*{0.5cm}
\end{center}
\caption{The parameter sets of two separable models of type 
Eq.~(\ref{schem}).}
\label{schempar}
\end{table}
The separable models are fitted to a neutron-proton $^1$S$_0$ 
scattering length, $a_{s} = -23.73 \, \text{fm}$, and effective 
range $r_{0} = 2.70 \, \text{fm}$.

\begin{figure}[t]
\vspace{-8mm}
\begin{center}
\includegraphics[scale=1.15,clip=]{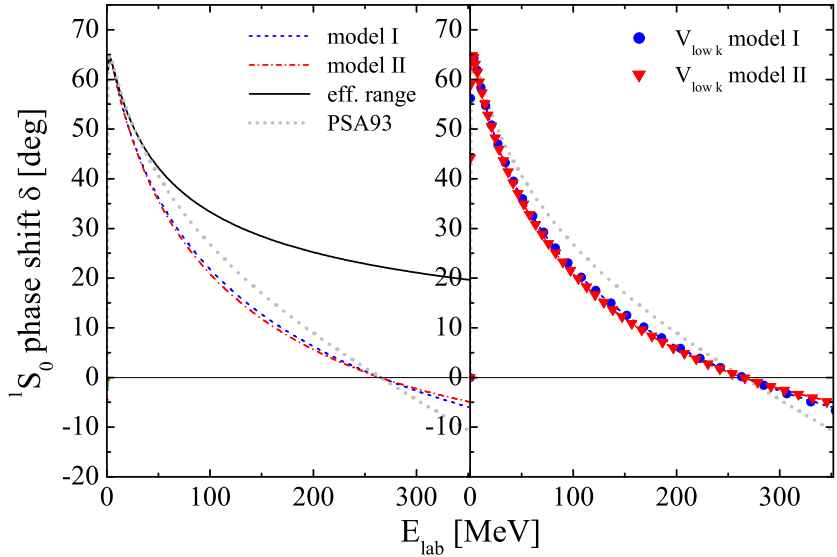}
\end{center}
\vspace{-8mm}
\caption{The left figure shows the phase shifts calculated
from the separable models I and II in comparison with the 
effective range expansion and the Nijmegen multi-energy phase shift 
analysis (PSA93)~\cite{PWA,NNonline}. In the right figure, the phase
shifts calculated from the corresponding low momentum interactions
are given. The low momentum interactions are obtained by solving 
the RG equation, Eq.~(\ref{schemrg}), to a scale $\la = 2.1 \,
\text{fm}^{-1}$.}
\label{schemps}
\end{figure}

In Fig.~\ref{schemps}, we show the phase shifts calculated from the
models of $V_{\text{bare}}$. We observe that the separable models I
and II can achieve a reasonably realistic description of the
nucleon-nucleon phase shifts for laboratory energies below
$E_{\text{lab}} \lesssim 350 \, \text{MeV}$, i.e., for relative 
momenta $k \lesssim 2.1 \, \text{fm}^{-1}$.

As discussed in the previous Section, the low momentum amplitudes
can be preserved with an effective interaction $V_{\text{low k}}$ 
acting in the Hilbert space of low momentum modes, $k < \la$, 
exclusively. The low momentum potential is renormalized by the high 
momentum modes according to an RG equation, of general form given 
by Eq.~(\ref{generalRGE}). In the current approach, we demand that 
the low momentum half-on-shell $T$ matrix (and consequently the phase 
shifts) are RG invariant, see also~\cite{Vlowk,Vlowkflow}. The RG equation 
which follows from this requirement is derived in~\cite{VlowkRG}. 
For separable interactions, the RG equation for the diagonal components 
of the low momentum interaction simplifies and is given in closed 
form by\footnote{The RG equation is regularized by imposing a small 
gap between the low momentum Hilbert space, $k \leqslant \la - 
\varepsilon$, and the cutoff $\la$ in the integration over intermediate 
momenta, $p_{\text{loop}} \leqslant \la$. In this way the $T$ matrix 
can be calculated to chosen accuracy. In other words, the effect of 
this regularization on the scattering amplitude is $\mathcal{O}(\varepsilon)$.}
\be
\frac{d}{d \Lambda} V_{\text{low k}} (k,k;\la) = \cfrac{2/\pi}{1-(k /
\Lambda)^{2}} \: \cfrac{V_{\text{low k}} (k,k;\la) \, V_{\text{low k}}
(\la,\la;\la)}{1 + \cfrac{2}{\pi} \, \mathcal{P} \int\limits_{0}^{\la}
\cfrac{V_{\text{low k}} (p,p;\la)}{p^{2}-\la^{2}} \, p^{2} dp } .
\label{schemrg}
\ee
The initial condition of the RG equation is given by the bare
interaction,
\be
V_{\text{low k}} (k,k;\la_0) = V_{\text{bare}} (k,k) ,
\ee
at an initial scale $\la_0$ where the high momentum modes are
negligible,
\be
V_{\text{bare}} (\la_0,\la_0) \approx 0 .
\ee

The RG equation is then used to evolve the bare potentials to low momenta,
thus removing the model-dependent realization of the high momentum
parts. We solve the separable RG equation to a cutoff $\la = 2.1 \,
\text{fm}^{-1}$, which corresponds to a laboratory energy scale up to 
which the realistic nucleon-nucleon interactions are constrained by 
experiment, $E_{\text{lab}} \sim 350 \, \text{MeV}$. First, we compare 
the phase shifts calculated in the low momentum space
\begin{align}
\tan\delta &= - k \: T_{\text{low k}}(k,k;k^2) \\
T_{\text{low k}}(k,k;k^2) &= \cfrac{V_{\text{low k}}(k,k;\la)}{1 +
\cfrac{2}{\pi} \, \mathcal{P} \int\limits_{0}^{\la} \cfrac{V_{\text{low k}}
(p,p;\la)}{p^{2}-k^{2}} \, p^{2} dp}
\end{align}
and verify in Fig.~\ref{schemps} that the phase shifts are indeed
reproduced by $V_{\text{low k}}$. In the RG approach, we have by 
construction
\be
T_{\text{low k}}(k',k;k^2) = T_{\text{bare}}(k',k;k^2) .
\ee

\begin{figure}[t]
\vspace{-8mm}
\begin{center}
\includegraphics[scale=1.15,clip=]{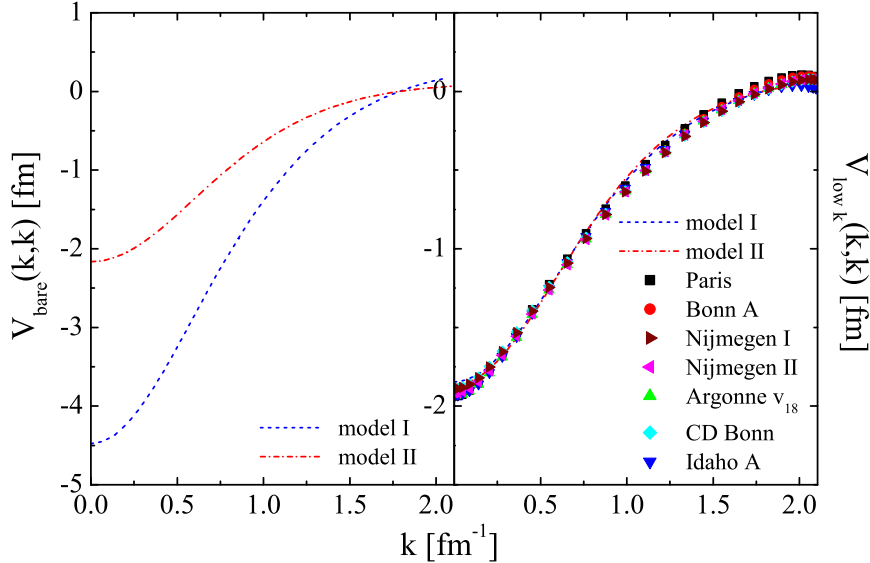}
\end{center}
\vspace{-8mm}
\caption{The diagonal momentum-space matrix elements of 
$V_{\text{bare}}$ and $V_{\text{low k}}$ derived from the
separable models I and II. As a comparison, we show the 
results for the diagonal matrix elements of the $V_{\text{low
k}}$ from Section~\ref{results} derived from the high precision 
nucleon-nucleon interactions.}
\label{schemdiag}
\end{figure}
In Fig.~\ref{schemdiag}, we show the resulting diagonal
matrix elements of $V_{\text{low k}}$ in comparison to the
$V_{\text{low k}}$ derived from the high precision nucleon-nucleon
interactions. Although the bare separable models I and II
differ substantially, we find that the low momentum interactions in
the truncated Hilbert space are identical. Furthermore,
we observe that, to a surprisingly good accuracy, these simple models
can in fact reproduce the diagonal matrix elements of the low momentum
interaction obtained from the realistic potential models.
In the RG equation for separable models, the matrix element
corresponding to zero phase shift remains invariant,
\be
V_{\text{low k}}(\kappa,\kappa;\la) = V_{\text{bare}}(\kappa,\kappa) =
T(\kappa,\kappa;\kappa^2)=0 .
\ee
Interestingly, we find that the diagonal matrix element of the
realistic $V_{\text{low k}}$ also changes sign approximately at
relative momentum $\kappa$.

\subsection{Organization}

This work is organized as follows. In the next Section we discuss the
calculational methods used for constructing the effective low momentum
interaction $V_{\text{low k}}$. In particular, we show how $\vlk$ is 
related to the solution of the Bloch-Horowitz equation. Since the
equivalence of the model space effective interaction methods and 
the RG approach is discussed in detail in~\cite{VlowkRG}, we will 
focus on the discussion of the effective interaction methods used 
in this work. In Section~\ref{results}, we present the results for 
the low momentum interactions derived from the realistic potential 
models. We give results for all partial waves with $J \leqslant 4$,
which provides ample evidence for the claimed model independence of
$\vlk$. By studying specific partial waves, we will show how the model 
indepence is reached as the cutoff is varied, and also demonstrate
its relation to the phase shift equivalence and the common
long-range pion physics of the different potential models. Finally, 
we verify that the low momentum interaction indeed reproduces the 
scattering data below $E_{\text{lab}} \lesssim 350 \, \text{MeV}$ and the 
deuteron binding energy with similar accuracy as the high precision 
potentials. After summarizing our results, we elaborate on the advantages of 
using $\vlk$ as the microscopic input interaction for nuclear many-body 
calculations.

\section{Calculational methods}

The schematic model provides a nice example illustrating the use of 
nonperturbative RG techniques to construct the low momentum
effective theory. One could therefore apply the 
same direct method as for the schematic model. However, in the general 
non-separable case, the direct solution of the RG equation  
would be more involved. Therefore, we have adapted
conventional effective interaction techniques, also referred to as 
model space methods, for the calculation of the low momentum 
interaction in the general case. These methods have been successfully
used to derive nuclear shell model interactions in a truncated Hilbert 
space, see e.g., the review~\cite{HJKO}. Both the RG and model
space techniques are concerned with integrating out the high-energy
states, such that the low-energy observables remain invariant in the
simpler effective theory. In spite of their similarities, the formal 
equivalence of conventional effective interaction theory and the RG 
was only recently demonstrated for the two-body
problem~\cite{VlowkRG}. There, we have shown that the low momentum 
Hamiltonian obtained from the solution of the RG equation is 
equivalent to the effective theory derived using Bloch-Horowitz or 
Lee-Suzuki projection methods. Since our current work makes use of 
this equivalence for the calculation of $V_{\text{low k}}$, we briefly 
discuss the relevant techniques below.

\subsection{Integrating out the high momentum modes by similarity
transformations} 

Consider a physical system, for which we are only interested in 
its low-energy properties. The first step in effective interaction
theory is to define Feshbach projection operators onto the physically 
important low-energy model space, the so-called $P$-space, and the 
high-energy complement, the $Q$-space,
\be
P = \sum_{i=1}^{d} | i \rangle \langle i | = \begin{pmatrix} 1 \: &
0 \\ 0 \: & 0 \end{pmatrix} \quad \text{and} \quad
Q = \sum_{i=d+1}^{\infty} | i \rangle \langle i | = \begin{pmatrix} 0 \: &
0 \\ 0 \: & 1 \end{pmatrix} .
\ee
The projection operators satisfy $P+Q=1$, $PQ=QP=0$,
$P^2=P$ and $Q^2=Q$, and we have written the projection operators
in matrix form as well. In most applications, the eigenstates of the 
unperturbed Hamiltonian $H_0$ are taken as basis states $| i \rangle$. 

The Schr\"odinger equation can then be written in block form as 
\be
\begin{pmatrix} PHP &PHQ \\ QHP & QHQ \end{pmatrix} \begin{pmatrix}
P \, | \Psi \rangle \\ Q \, | \Psi \rangle \end{pmatrix} = E \begin{pmatrix}
P \, | \Psi \rangle \\ Q \, | \Psi \rangle \end{pmatrix} .
\label{blockeq}
\end{equation}
By using the second block-row of Eq.~(\ref{blockeq}), the $Q$-space 
projection $Q \, | \Psi \rangle$ can be eliminated and one obtains 
the fundamental equations of Bloch-Horowitz effective interaction 
theory~\cite{Bloch,BlochHoro}
\be
P \mathcal{H}_{\text{low k}}^{\text{BH}}(E) P \, | \Psi \rangle = E P
\, | \Psi \rangle ,
\ee
where the effective Hamiltonian $\mathcal{H}_{\text{low k}}^{\text{BH}}(E)$ is 
obtained by the solution of the Bloch-Horowitz equation
\be
\mathcal{H}_{\text{low k}}^{\text{BH}}(E) = H + H \: \frac{1}{E-QHQ} \: H .
\ee
The Bloch-Horowitz equation generates an effective Hamiltonian, whose
$P$-space projection $H_{\text{low k}}^{\text{BH}}(E) = 
P \, \mathcal{H}_{\text{low k}}^{\text{BH}}(E) \, P$ is operative 
only within the low-energy model space and, e.g., enables an exact 
diagonalization in shell model applications. The $Q$-space states 
have been decoupled from the problem in a way that preserves the 
low-energy spectrum exactly, i.e., the $d$ lowest-lying eigenvalues. 
The eigenstates of $H_{\text{low k}}^{\text{BH}}(E)$ are simply 
given by the $P$-space projections of the exact 
eigenstates $P \, | \Psi_n \rangle$. The effective Hamiltonian 
$H_{\text{low k}}^{\text{BH}}(E)$ depends on the exact energy eigenvalue
one is solving for, and thus necessitates a self-consistent treatment. 
Although the self-consistency problem can be solved nicely using the Lanczos 
algorithm~\cite{Haxton1}, the use of energy-dependent
two-body interactions in many-body calculations can pose computational 
and conceptual difficulties.

A refinement of the Bloch-Horowitz theory that avoids these
difficulties is the Lee-Suzuki similarity
transformation~\cite{LS1,LS2}. The Lee-Suzuki method constructs 
a similarity transformation that brings the Hamiltonian to the 
following block structure
\be
\Theta^{-1} \, H \, \Theta = \mathcal{H}_{\text{low k}}^{\text{LS}} = 
\begin{pmatrix} P \, \mathcal{H} \, P && P \, \mathcal{H} \, Q \\ 0 && Q
\, \mathcal{H} \, Q \end{pmatrix} .
\label{block}
\ee
Since the determinant of a block-triangular matrix factorizes into the
determinants of the independent $P$-space and $Q$-space blocks, the 
low-energy spectrum can be calculated exactly from the smaller effective 
model space Hamiltonian $H_{\text{low k}}^{\text{LS}} = P \,
\mathcal{H}_{\text{low k}}^{\text{LS}} \, P$. As in the Bloch-Horowitz 
approach, the eigenstates of $H_{\text{low k}}^{\text{LS}}$ are given
by $P \, | \Psi_n \rangle$. Making an ansatz for the Lee-Suzuki
similarity transformation, one parametrizes $\Theta$ as a
non-orthogonal transformation in terms of the so-called wave operator
$\omega$
\be
\Theta = 1+\omega = \begin{pmatrix} 1 \: & 0 \\ \omega \: & 1 \end{pmatrix}
\quad \text{and} \quad \Theta^{-1} = 1-\omega = \begin{pmatrix} 1 \: & 0 \\
- \omega \: & 1 \end{pmatrix} ,
\ee
where $\omega = Q \, \omega \, P$ connects the $Q$ and
$P$-space. Inserting the ansatz into Eq.~(\ref{block}) leads to a 
non-linear constraint on $\omega$, the decoupling equation,
\be
Q \, \mathcal{H}_{\text{low k}}^{\text{LS}} \, P = 0 \quad \Longrightarrow
\quad Q H P + Q H Q \, \omega - \omega \, P H P - \omega \, P H Q \,
\omega = 0 . 
\label{decouple}
\ee
Once a solution to this non-linear operator equation is
found, the energy-independent effective Hamiltonian is given by 
\be
H_{\text{low k}}^{\text{LS}} = P H P + P H Q \, \omega \, P = P H P + 
P V Q \, \omega \, P ,
\label{hlowk}
\ee
where we have assumed that $P$ and $Q$ commute with the unperturbed 
Hamiltonian. 

The relation to the Bloch-Horowitz theory can be clarified by 
inserting the Lee-Suzuki solution, Eq.~(\ref{hlowk}), into the
decoupling equation, Eq.~(\ref{decouple}), and solving for the wave
operator. The resulting formal expression for the wave operator allows one to 
express the energy-independent Lee-Suzuki $H_{\text{low
k}}^{\text{LS}}$ as a summation of the lowest $d$, self-consistent 
Bloch-Horowitz solutions $H_{\text{low k}}^{\text{BH}}(E)$,
where each term is weighted by the projection of the corresponding eigenstate 
of the effective theory $P \, | \Psi_n \rangle \widetilde{ \bigl( \langle \Psi_n | \,
P \bigr)}$\footnote{The tilde denotes the bi-orthogonal complement, 
since projections of orthogonal $| \Psi_n \rangle$ are not necessarily 
orthogonal.}
\begin{align}
H_{\text{low k}}^{\text{LS}} &= \sum_{n=1}^{d} \biggl( P H P + P H \,
\frac{1}{E_n-QHQ} \, H P \biggr) \: P \, | \Psi_n \rangle \widetilde{
\bigl( \langle \Psi_n | \, P \bigr)} \nonumber \\[2mm]
&= \sum_{n=1}^{d} H_{\text{low k}}^{\text{BH}}(E_n) \: P \, | \Psi_n
\rangle \widetilde{\bigl( \langle \Psi_n | \, P \bigr)} .
\label{energyavg}
\end{align}

In this work, we apply the Lee-Suzuki formalism to the free-space 
nucleon-nucleon problem to derive the energy-independent low momentum
interaction 
\be
V_{\text{low k}} = H_{\text{low k}}^{\text{LS}} - P H_0 P = P V P + 
P V Q \, \omega \, P .
\ee
This approach has been proven to be equivalent to the solution of the 
RG equation that results from imposing a momentum-space cutoff and 
demanding RG invariant half-on-shell (HOS) $T$ matrices~\cite{VlowkRG}. 
For a given partial wave, the $P$ and $Q$ operators are defined in a 
continuous plane wave basis as
\be
P = \frac{2}{\pi} \,  \int_{0}^{\Lambda} p^{2} dp \, | p \rangle
\langle p | \quad \text{and} \quad Q = \frac{2}{\pi} \, 
\int_{\Lambda}^{\infty} q^{2} dq\, | q \rangle \langle q | .
\ee

There are several methods to solve the non-linear decoupling
equation for the wave operator. For the two-body problem, e.g., the exact
solutions for the eigenstates can be used, and $V_{\text{low k}}$ can
be calculated directly using Eq.~(\ref{energyavg}). For more complex 
applications, one solves the decoupling equation using iterative
techniques. Since the iterative methods are numerically very robust,
we have utilized the iterative algorithm of Andreozzi~\cite{Andreozzi} 
to calculate $V_{\text{low k}}$. The defining equations of this 
iteration scheme are given by
\begin{align}
\mathcal{X}_{0} &= - \bigl( {QHQ} \bigr)^{-1} \, QHP \\[1mm]
\mathcal{X}_{n} &= \bigl( \mathcal{Q}(\omega_{n-1}) \bigr)^{-1} \,
\mathcal{X}_{n-1} \, H_{\text{low k}}^{\text{LS}}(\omega_{n-1}) , \quad
\text{for} \:\: n \geqslant 1 ,
\end{align}
where
\begin{align}
\omega_{n} &= \mathcal{X}_{0} + \mathcal{X}_{1} + \ldots +
\mathcal{X}_{n} \\[1mm]
\mathcal{Q}(\omega_{n}) &= QHQ - Q \, \omega_{n} \, PHQ
\end{align}
and
\be
H_{\text{low k}}^{\text{LS}}(\omega_{n}) = PHP + PHQ \,
\omega_{n} \, P .
\label{ALS}
\ee
The iteration converges when $H_{\text{low k}}^{\text{LS}}(\omega_{n})
\approx H_{\text{low k}}^{\text{LS}}(\omega_{n-1})$.

A convenient feature of the algorithm is that each iteration 
only requires one matrix inversion in the $Q$-space. The $Q$-space
matrix inversion is manageable, typically requiring a set of $\approx 60$
Gauss-Legendre mesh points extending to $\approx 25 \,\text{fm}^{-1}$
to include the high momentum behavior of the input interaction.
Finally, we mention that variations of the algorithm 
exist in which all matrix inversions occur in $P$-space, which can 
offer significant computational advantages for large-scale
problems~\cite{Andreozzi}.

\subsection{A refinement: hermitian low momentum interactions}
\label{hermitian}

By construction, the Lee-Suzuki $H_{\text{low k}}^{\text{LS}}$ 
is non-hermitian because the eigenstates of the effective theory 
are projections of the full eigenstates onto the low momentum subspace. 
The non-hermiticity is a result of eliminating the energy-dependence 
of the Bloch-Horowitz theory by means of the non-orthogonal similarity
transformation.\footnote{Although the Bloch-Horowitz Hamiltonian 
$H_{\text{low k}}^{\text{BH}}(E_n)$ is hermitian for a given energy 
$E_n$, one encounters the same non-orthogonality problem between
eigenstates, since the Hamiltonian changes self-consistently depending
on the energy $E_n$.} However, one can perform an additional
similarity transformation on $H_{\text{low k}}^{\text{LS}}$ 
to obtain an energy-independent and hermitian
Hamiltonian~\cite{herm1,herm2}, which we denote by 
$\overline{H}_{\text{low k}}^{\text{LS}}$ and correspondingly, the 
hermitian low momentum interaction by $\overline{V}_{\text{low k}}$.
Both similarity transformations combine to give a
unitary transformation, which is equivalent to the
well-known Okubo transformation~\cite{Okubo}.

Here, we follow the method of Andreozzi~\cite{Andreozzi} in
constructing the second similarity transformation. Using the 
definition of the Lee-Suzuki transformation, Eq.~(\ref{block}), 
one finds
\be
\Theta^{\dagger} \Theta \, \mathcal{H}_{\text{low k}}^{\text{LS}} =
\bigl(\mathcal{H}_{\text{low k}}^{\text{LS}}\bigr)^{\dagger} \, 
\Theta^{\dagger} \Theta ,
\label{andreherm1}
\ee
where we have used $H = H^\dagger$. Projecting onto the low momentum 
space, the $P$-space block can be written as
\be
( \, P + \omega^{\dagger} \, \omega \, ) \, H_{\text{low k}}^{\text{LS}} =
\bigl(H_{\text{low k}}^{\text{LS}}\bigr)^{\dagger} \, 
( \, P + \omega^{\dagger} \, \omega \, ) .
\label{andreherm2}
\ee
The operator $P + \omega^{\dagger} \, \omega$ is hermitian 
and positive definite. Therefore, it can be written by means of a 
Cholesky decomposition as a product of a lower-triangular matrix 
$L$ and its adjoint $L^{\dagger}$. Writing
\be
( \, P + \omega^{\dagger} \, \omega \, ) = L L^{\dagger}
\ee 
in Eq.~(\ref{andreherm2}) and multiplication by $L^{-1}$ from the 
left and $(L^{\dagger})^{-1}$ from the right, one finds
\be
L^{-1} \, \bigl(\mathcal{H}_{\text{low k}}^{\text{LS}}\bigr)^{\dagger} \,
L = L^{\dagger} \, \mathcal{H}_{\text{low k}}^{\text{LS}} \,
(L^{\dagger})^{-1}= \bigl( L^{-1} \, \bigl(\mathcal{H}_{\text{low
k}}^{\text{LS}}\bigr)^{\dagger} \, L \bigr)^{\dagger} . 
\label{andreherm3}
\ee
Therefore, we can write the hermitian 
$\overline{H}_{\text{low k}}^{\text{LS}}$ directly in terms of the
Lee-Suzuki transformation and the Cholesky decomposition as
\begin{equation}
\overline{H}_{\text{low k}}^{\text{LS}} = L^{\dagger} \, H_{\text{low
k}}^{\text{LS}} \, (L^{\dagger})^{-1} ,
\label{andreherm4}
\end{equation}
and the combined transformation $U = \Theta \, (L^{\dagger})^{-1}$ is
unitary. 

We mention that the differences between the non-hermitian
$V_{\text{low k}}$ and the hermitian $\overline{V}_{\text{low k}}$ 
can be understood in the RG approach. The non-hermitian effective 
theory is obtained by introducing a cutoff in momentum space and 
imposing RG invariance for the HOS $T$ matrix with an
energy-independent low momentum potential. Since the initial conditions 
of the RG equation are given by the bare interaction, this implies 
that the effective theory preserves the low momentum components of
the full wave functions, in addition to the phase shifts and the bound
state poles. The choice of HOS $T$ matrix invariance is motivated 
by the physical observation that the low momentum components of the
wave functions mainly probe the long-range OPE part of the
interaction. Conversely, the hermitian effective theory results from
RG invariant on-shell $T$ matrices. Thus, the resulting hermitian
$\overline{V}_{\text{low k}}$ no longer preserves the low momentum 
components of the full wave functions, although the observable 
phase shifts and bound state poles are preserved. In practice, we
will find that the numerical differences between the hermitian and 
non-hermitian interactions are quite small.

Finally, we note that the Lee-Suzuki similarity transformation 
method is an algebraic realization of the more general folded 
diagram method of Kuo, Lee and Ratcliff~\cite{HJKO,KLR,KuoOsnes}. 
In this connection, one can recast the results of this Section in a 
diagrammatic formalism, which can be useful in many-body 
systems. Moreover, the folded diagram formulation has
been applied to the two-body problem to prove 
diagrammatically that the low momentum effective theory 
preserves the low-energy scattering observables of the input
interaction~\cite{Vlowk}. 

\section{Results}
\label{results}

We now discuss the main results for the low momentum interaction $\vlk$, 
which we have obtained by integrating out the high momentum components 
of different realistic potential models.
Most of the results shown here are for a momentum cutoff of 
$\Lambda = 2.1 \, \text{fm}^{-1}$, which corresponds to a 
distance scale $1/\Lambda \sim 0.5 \,\text{fm}$. 
With cutoffs chosen in this range, the model-dependent 
parameterization of the repulsive core at $r \lesssim 0.5 \,
\text{fm}$ in the bare interactions is not resolved in the 
effective theory. Consequently, one is able to reproduce 
the deuteron pole and scattering observables below 
$E_{\text{lab}} = 350 \, \text{MeV}$ without introducing 
a specific short-distance model. Moreover, the RG invariance 
of the decimation guarantees that our conclusions are independent  
of the precise value of the cutoff.

\begin{figure}[t]
\vspace{-8mm}
\begin{center}
\includegraphics[scale=1.06,clip=]{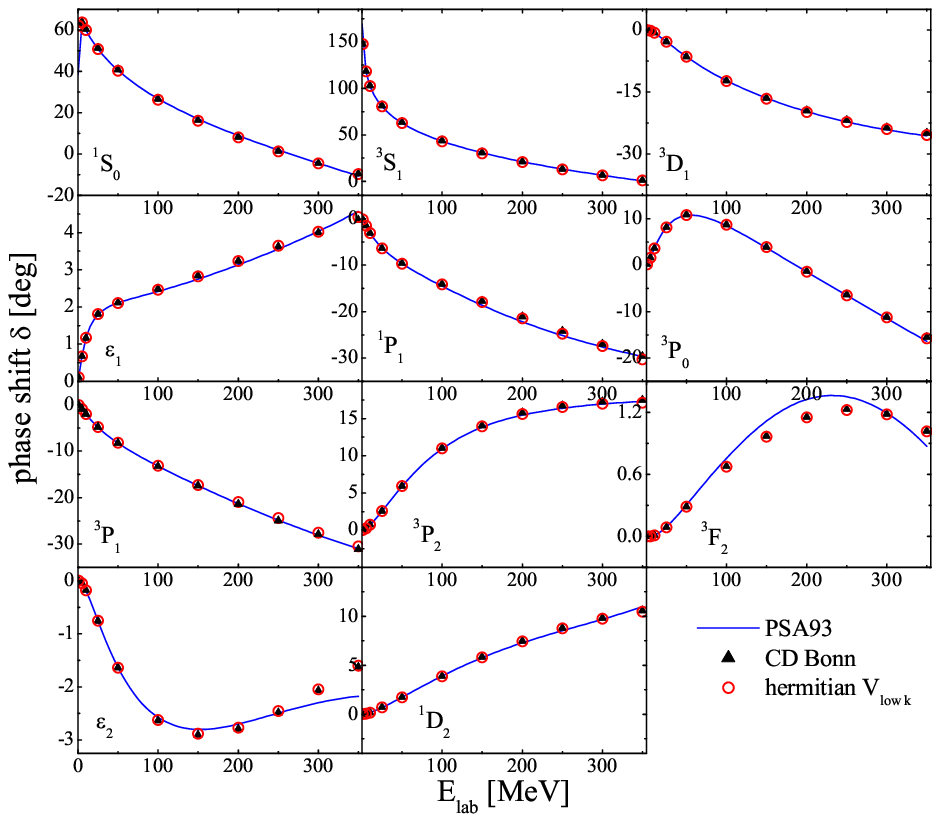}
\end{center}
\vspace{-8mm}
\caption{The neutron-proton phase shifts and mixing parameters
calculated from the hermitian $\overline{V}_{\text{low k}}$ in the 
partial waves $J \leqslant 4$ are shown. The $\overline{V}_{\text{low k}}$
is derived from the CD Bonn potential with a cutoff of 
$\Lambda = 2.1 \, \text{fm}^{-1}$. In comparison, the phase shifts 
and mixing parameters calculated from the bare CD Bonn potential as 
well as the results of the Nijmegen multi-energy phase shift 
analysis~\cite{PWA,NNonline} are included.}
\label{vlowkps1}
\vspace{-6mm}
\begin{center}
\includegraphics[scale=1.06,clip=]{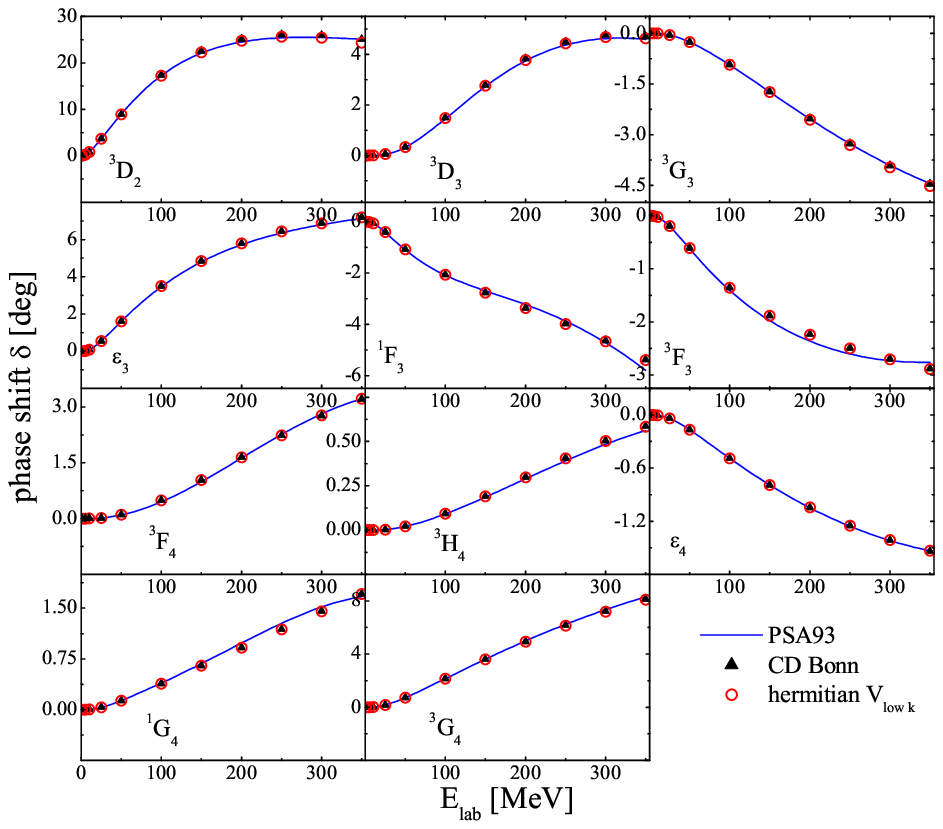}
\end{center}
\vspace{-8mm}
\caption{The neutron-proton phase shifts and mixing parameters are
continued from Fig.~\ref{vlowkps1}, for details we refer to the text
of Fig.~\ref{vlowkps1}.}
\label{vlowkps2}
\end{figure}

Before we present our results in detail, we summarize our main findings
and refer the reader to the key figures. The most important properties of
$\vlk$ are:
\begin{enumerate}
\item $\vlk$ renormalizes to a
nearly universal interaction as the cutoff is lowered 
to the scale of $\Lambda \sim 2.1 \, \text{fm}^{-1}$. We will argue that 
the model-independence of the diagonal 
matrix elements of $\vlk$, as shown in
Fig.~\ref{vlowkdiag1} and Fig.~\ref{vlowkdiag2}, is driven by the phase
shift equivalence of the input models. The correlation between 
the phase shift equivalence and the collapse of the low momentum 
interactions is clearly demonstrated in Fig.~\ref{vlowkpseq}. 
We also find practically identical results for the hermitian $\vlkb$, 
see Fig.~\ref{hermdiag1} and Fig.~\ref{hermdiag2}.\\
\item $\vlk$ reproduces the phase shifts and the deuteron binding
energy with similar accuracy as the high precision models. The phase
shifts calculated from $\vlk$ are shown in Fig.~\ref{vlowkps1} and
Fig.~\ref{vlowkps2}.\\
\item In the effective theory one does not have to make model-dependent 
assumptions on the short-distance interaction. The 
RG evolution filters out the short-distance details of 
the input potentials, while preserving the model-independent effects
of the high momentum modes on the low momentum observables, and leaving 
the long-range part (OPE) of the interaction intact. This can be seen most
clearly in Fig.~\ref{collapse1s0}, Fig.~\ref{collapse3s1} and 
Fig.~\ref{collapse3sd1}.
\end{enumerate}
We also point out that the numerical differences between $\vlk$ and the
hermitian $\vlkb$ are quite small, see Fig.~\ref{hermcomp}. 
For brevity, we mostly present results for $\vlk$ in this work, since
the $\vlkb$ is easily derived from $\vlk$ using the transformation described
in Section~\ref{hermitian}.

\subsection{Model-independence of the low momentum interaction}

For cutoffs around $\Lambda \sim 2.1 \text{fm}^{-1}$, the effective 
theory preserves the nucleon-nucleon phase shifts up to laboratory
energies of $E_{\text{lab}} \sim 350 \, \text{MeV}$. This corresponds 
to the scale of the scattering data that all the high precision 
potential models are fitted to. We have argued that, based on the
separation of mass scales, the model-dependence of the realistic
interactions should be reduced as we integerate out the ambiguous 
high momentum modes.
\begin{figure}[t]
\vspace{-8mm}
\begin{center}
\includegraphics[scale=1.145,clip=]{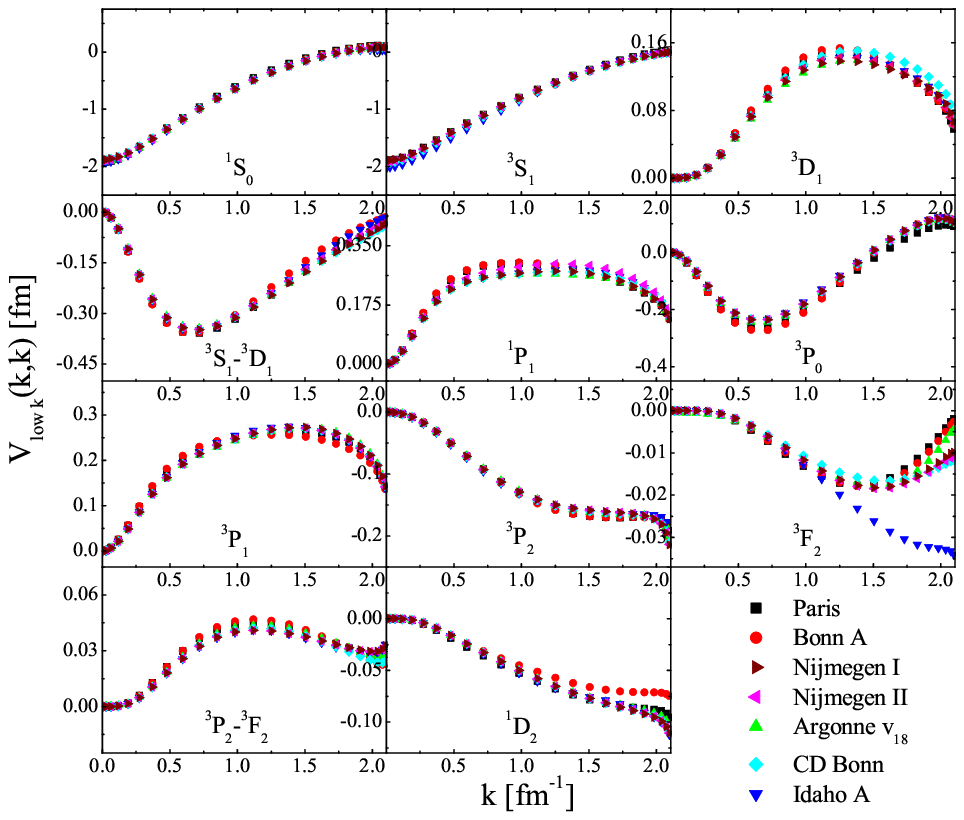}
\end{center}
\vspace{-8mm}
\caption{Diagonal momentum-space matrix elements of the $\vlk$
obtained from the different potential models for a cutoff 
$\Lambda = 2.1 \, \text{fm}^{-1}$. Results are shown for the 
partial waves $J \leqslant 4$.}
\label{vlowkdiag1}
\vspace{-6mm}
\begin{center}
\includegraphics[scale=1.145,clip=]{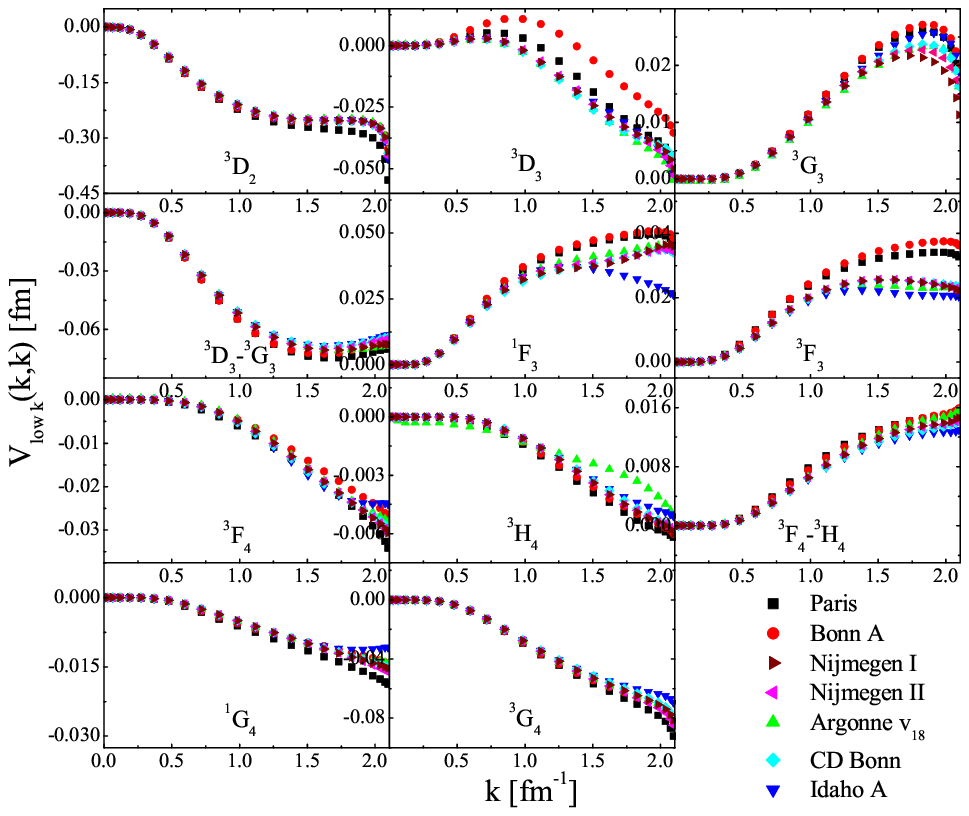}
\end{center}
\vspace{-8mm}
\caption{The diagonal matrix elements of $\vlk$ are continued from
Fig.~\ref{vlowkdiag1}.}
\label{vlowkdiag2}
\end{figure}
In Fig.~\ref{vlowkdiag1} and Fig.~\ref{vlowkdiag2}, we present the
results of the RG decimation. We observe that the diagonal matrix 
elements of the low momentum interactions derived from the different 
potential models are nearly identical, and thus insensitive to the
input potential model for cutoffs $\Lambda \lesssim 2.1 \, \text{fm}^{-1}$.
The largest relative differences between the $\vlk$ are
found in the $^3$F$_2$, $^3$D$_3$, $^1$F$_3$ and $^3$F$_3$ partial
waves. These differences are correlated with the deviations in the 
phase shifts, i.e., the different accuracies, of the potential models. 
We will discuss this point in detail below.

We note that if we perform the additional similarity transformation 
to obtain the resulting hermitian $\vlkb$, we find the same 
model-independence and practically identical diagonal matrix 
elements, as shown in Fig.~\ref{hermdiag1} and Fig.~\ref{hermdiag2} 
in the Appendix. Since the hermitian $\vlkb$ is obtained by 
symmetrizing $\vlk$, it is expected that the effects on the 
diagonal matrix elements are not significant.

\begin{figure}[t]
\vspace{-8mm}
\begin{center}
\includegraphics[scale=1.15,clip=]{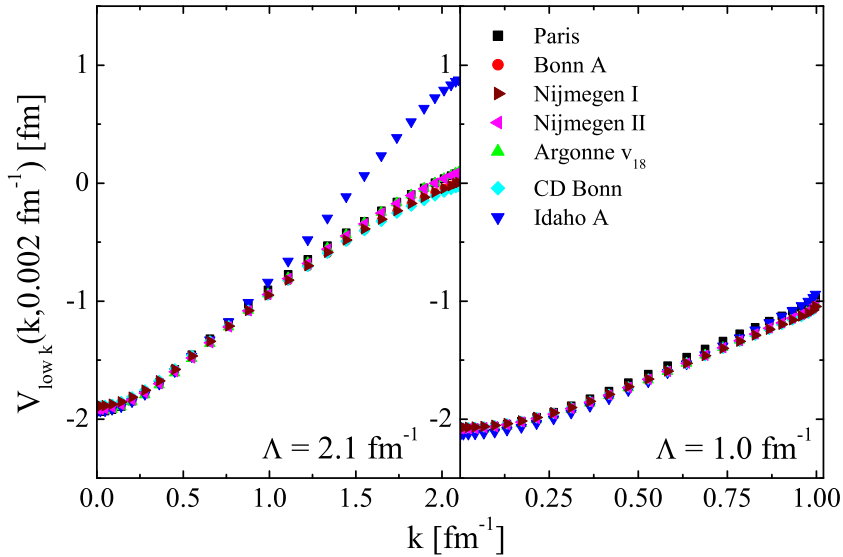}
\end{center}
\vspace{-8mm}
\caption{The off-diagonal momentum space matrix elements of the $\vlk$
obtained for a cutoff $\Lambda = 2.1 \, \text{fm}^{-1}$ (left) and $\Lambda =
1.0 \, \text{fm}^{-1}$ (right) in the $^1$S$_0$ partial wave.}
\label{offdiag1s0}
\end{figure}
One obtains further insight into the collapse of $\vlk$ by studying 
the off-diagonal matrix elements. These are shown in the $^1$S$_0$
partial wave in Fig.~\ref{offdiag1s0}. We find a similar nearly
universal behavior for the off-diagonal matrix elements of $\vlk$, 
although those calculated from the Idaho potential differ at 
$k' \approx 1.2 \, \text{fm}^{-1}$, which is approximately the 
mass scale given by the $2 \pi$ exchange. The conventional models 
include $2 \pi$ contributions in which non-nucleonic 
intermediate states are explicitly taken into account. In contrast, the Idaho 
potential starts from a local, effective Lagrangian, in which nucleons 
and pions are the only explicit degrees of freedom. For example,
contributions from $\Delta$ intermediate states between two 
successive pion exchanges are contracted to a renormalized 
${\mathcal L}^{\text{eff}}_{\pi \text{N} \text{N}}$ vertex. Moreover, the 
conventional potential models all contain a strong, short-range repulsion 
associated with the exchange of heavy mesons, mostly originating from
the $\omega$ exchange. In the Idaho potential, the repulsive core
is not explicitly present. However, the effects of the short-range 
repulsion on low-energy observables are contained in contact terms 
that are fitted to the low-energy data~\cite{Idaho}. Therefore, in 
momentum space, the Idaho potential changes over to contact terms 
at momentum transfers above the range given by $2 m_\pi$. 

\begin{figure}[t]
\vspace{-8mm}
\begin{center}
\includegraphics[scale=0.95,clip=]{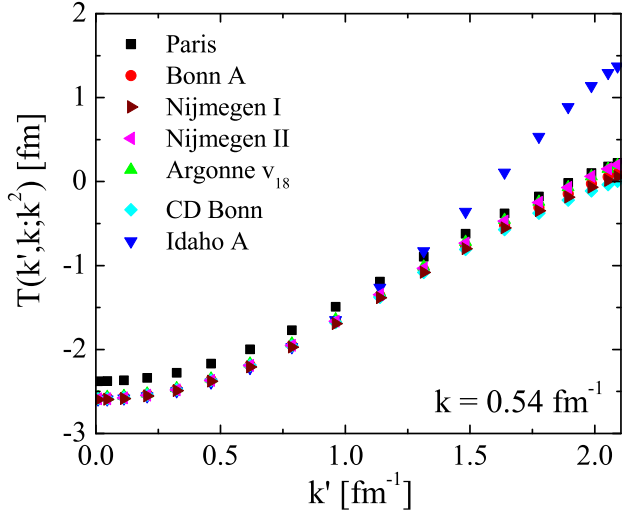}
\end{center}
\vspace{-8mm}
\caption{Comparison of the HOS $T$ matrices calculated from the
different potential models in the $^1$S$_0$ partial wave.}
\label{HOS1s0}
\end{figure}
While the Idaho potential is phase shift equivalent to the 
conventional models, the HOS $T$ matrices, i.e., the low momentum
components of the wave functions, differ significantly from the 
conventional models at energies and momenta larger than the above
scale of $2 m_\pi$. In Fig.~\ref{HOS1s0}, we find that the 
conventional models give remarkably similar HOS $T$ matrices 
over their range of phase shift equivalence. This approximate 
HOS $T$ matrix equivalence can be understood by expanding the 
HOS $T$ matrix around the fully on-shell part. The difference can be
written as a wound integral that contains the off-shell behavior, 
\be
T(k',k;k^2) = T(k,k;k^2) + \frac{k^{2} - k'^{2}}{k'} \int_{0}^{\infty}
dr \: \sin({k'r}) \: \bigl( u_{k}(r) - v_{k}(r) \bigr) , 
\label{hosidentity}
\ee
where $u_{k}(r) = r \, \Psi_{k}(r)$ denotes the exact and
$v_{k}(r)$ is the asymptotic S-wave wave function
\be 
\lim\limits_{r \rightarrow \infty} u_{k}(r) \rightarrow v_{k}(r) \sim
\sin({kr}+\delta) .
\ee

We can use Eq.~(\ref{hosidentity}) to express the difference of the HOS
$T$ matrices for two different, but phase shift equivalent bare
interactions. This leads to
\be 
\Delta T(k',k;k^2) = \frac{k^{2}-k'^{2}}{k'} \int_{0}^{R} dr \:
\sin({k'r}) \: \Delta u_{k}(r) ,
\ee
where $R$ is the distance over which the two models differ
substantially at short distances, which we can take for 
simplicity as the size of the repulsive core. Regardless of 
the details of the differences, at scattering energies well below the
strength of the short-range repulsion, the differences between
the wave functions in the core are suppressed by strong 
two-body correlations. Therefore, it is physically reasonable that 
the conventional realistic potentials give similar HOS $T$ 
matrices at low energies. 

The deviations in the off-diagonal matrix elements of $\vlk$ derived
from the Idaho potential are obviously removed, if we integrate out
further below $2 m_\pi$. As shown in Fig.~\ref{offdiag1s0}, we find that 
the off-diagonal matrix elements in the $^1$S$_0$ partial wave
collapse onto one curve as well for $\Lambda \lesssim 1.2 \,
\text{fm}^{-1}$. 

\begin{figure}[t]
\vspace{-6mm}
\begin{center}
\includegraphics[scale=1.15,clip=]{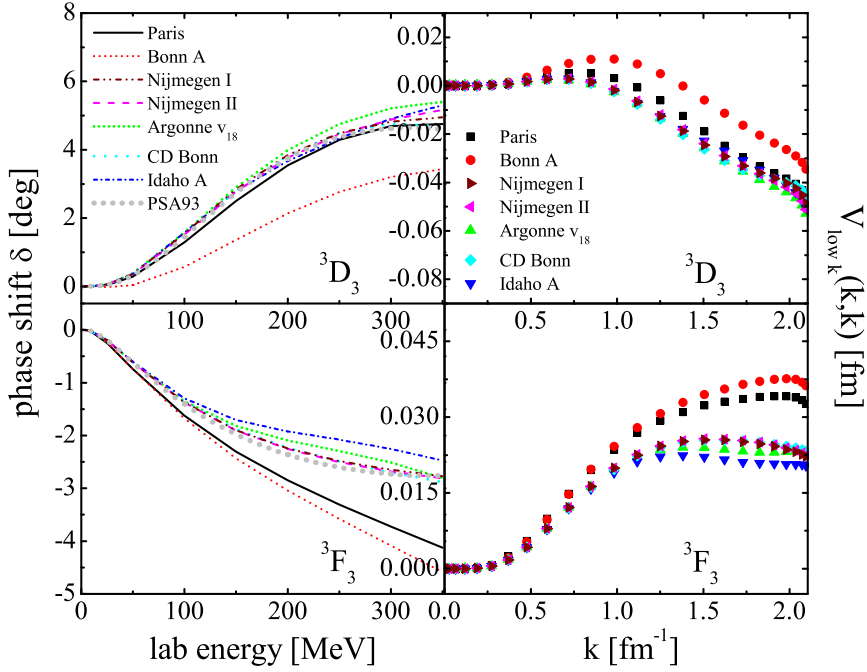}
\end{center}
\vspace{-6mm}
\caption{Comparison of the phase shifts given by the $V_{\text{NN}}$
models and the diagonal matrix elements of $\vlk$ in the $^3$D$_3$ and
$^3$F$_3$ partial waves. Results for $\vlk$ are shown with a cutoff
$\Lambda = 2.1 \, \text{fm}^{-1}$.}
\label{vlowkpseq}
\end{figure}
The preceding observations strongly suggest that the collapse of 
the diagonal matrix elements of $\vlk$ is driven by the phase 
shift equivalence of the input models, while the collapse of the 
off-diagonal matrix elements is controlled by approximate 
HOS $T$ matrix equivalence at low energy and momentum. The latter 
can in turn be mainly 
attributed to a common off-shell behaviour in the OPE interaction,
with some differences arising from the two-pion exchange and the
crossover to contact interactions. The correlation between the
diagonal matrix elements and the phase shifts is nicely shown in the
$^3$D$_3$ and $^3$F$_3$ partial waves, where we observe the largest
relative deviations from model-independence. The left panels of 
Fig.~\ref{vlowkpseq} clearly show that the older generation 
Paris and Bonn potential models give significantly different
phase shifts in these partial waves. This leads to the observed 
differences in the $\vlk$ derived from these bare interactions, 
as can be seen in the right panels of Fig.~\ref{vlowkpseq}. For comparison,
we have also shown the results of the Nijmegen multi-energy phase shift 
analysis (PSA93)~\cite{PWA,NNonline}.

\begin{figure}[t]
\vspace{-8mm}
\begin{center}
\includegraphics[scale=0.95,clip=]{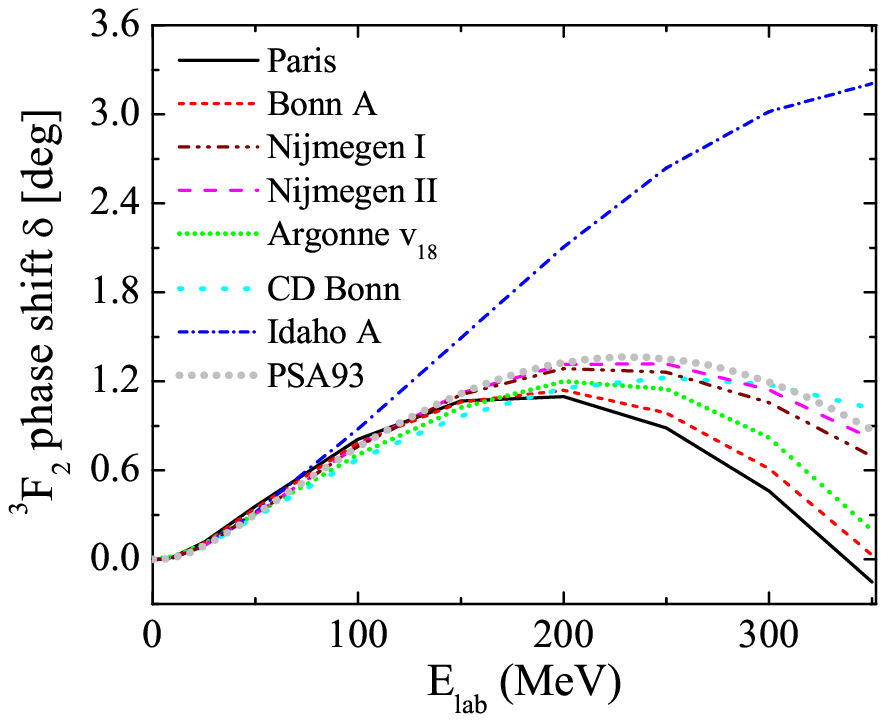}
\end{center}
\vspace{-8mm}
\caption{Phase shifts in the $^3$F$_2$ partial wave calculated from 
different $V_{\text{NN}}$ models.}
\label{fwavephase}
\vspace{-8mm}
\begin{center}
\includegraphics[scale=1.15,clip=]{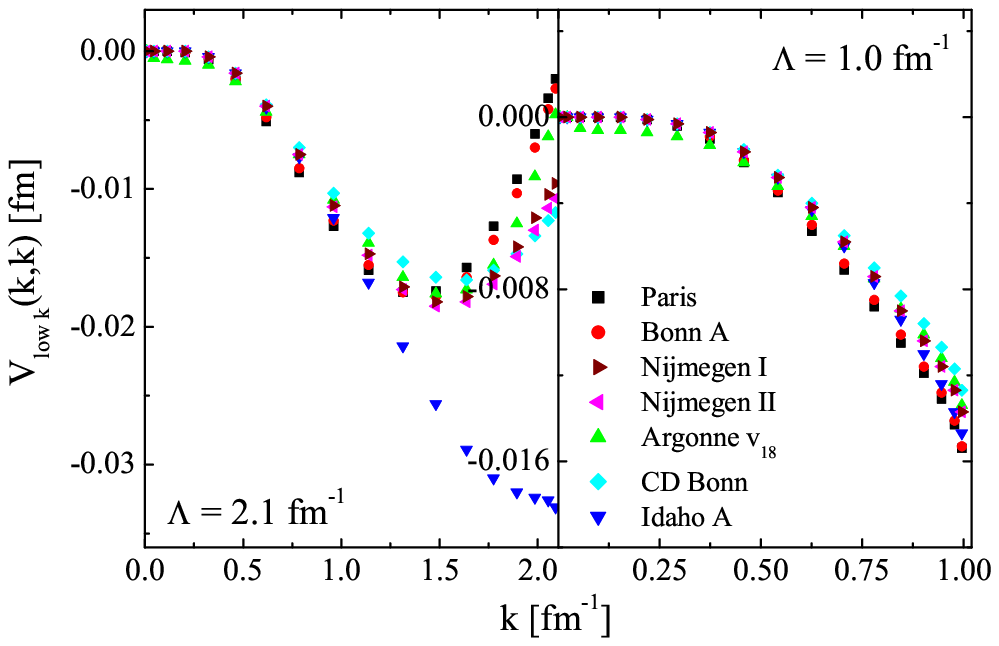}
\end{center}
\vspace{-8mm}
\caption{The diagonal matrix elements of $\vlk$ for $\Lambda=2.1 \,
\text{fm}^{-1}$ (left) and $\Lambda = 1.0 \, \text{fm}^{-1}$ (right)
in the $^3$F$_2$ partial wave.}
\label{fwavediag}
\end{figure}
A further example of this correlation can be seen in
Fig.~\ref{fwavephase}, where the Idaho potential leads 
to considerably different phase shifts in the $^3$F$_2$
partial wave above laboratory energies $E_{\text{lab}} \gtrsim 100 \,
\text{MeV}$. This corresponds to a relative momentum of $k \gtrsim 1.0
\, \text{fm}^{-1}$. Accordingly in Fig.~\ref{fwavediag}, we find 
that the diagonal matrix elements of $\vlk$ derived from the 
Iadaho potential deviate from the others for momenta larger than 
this scale. As for the off-diagonal matrix elements in the
$^1$S$_0$ partial wave, these differences vanish as the cutoff 
is lowered to $\Lambda \lesssim 1.0 \, \text{fm}^{-1}$. 

In the N$^2$LO version of the Idaho potential used here,
the contact interactions do not contribute to the $L
\geqslant 3$ partial waves. Thus, in the $^3$F$_2$ partial wave the
Idaho potential is completely given by pion exchange. Our results
demonstrate that the effects of the short-distance physics cannot
simply be ignored. Rather, their renormalization effects on the low 
momentum physics has to be taken into account. If we lower the cutoff
to $\Lambda \lesssim 1.0 \, \text{fm}^{-1}$, the effects of the high
momentum modes is sufficiently small, and one finds that indeed the
physics is given adequately by pion exchange. 
This can be understood in the Born approximation, which is a 
reasonable approximation for the high partial waves, especially at 
low energies. In the Born approximation, the integral term of the
Lippmann-Schinger equation is neglected, and thus the low momentum
interaction in the cutoff Hilbert space is given by simply
cutting off the bare interaction.

\begin{figure}[t]
\vspace{-8mm}
\begin{center}
\includegraphics[scale=1.15,clip=]{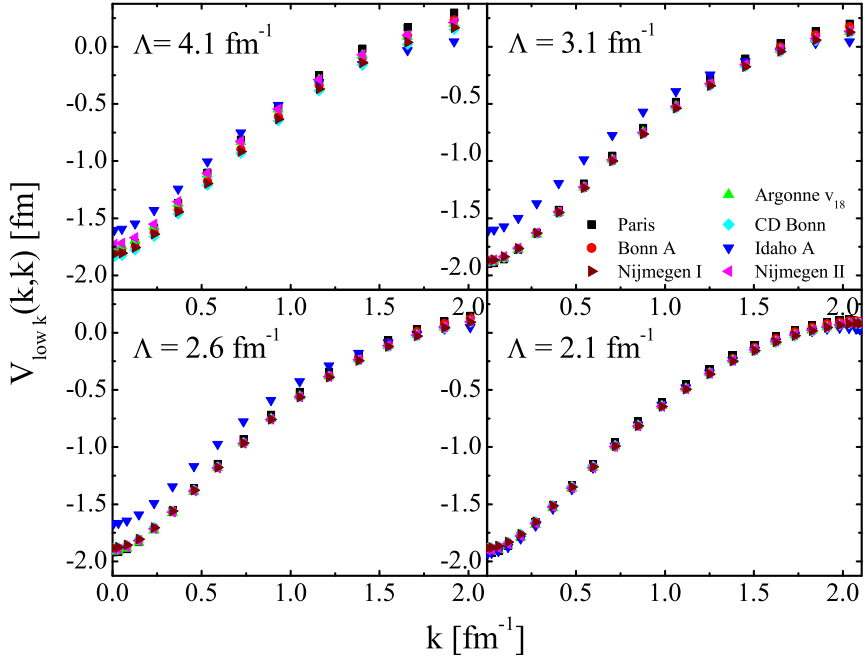}
\end{center}
\vspace{-8mm}
\caption{The collapse of the diagonal momentum-space matrix elements
of $\vlk$ as the cutoff is lowered to $\Lambda = 2.1 \, \text{fm}^{-1}$ 
in the $^1$S$_0$ partial wave.}
\label{collapse1s0}
\end{figure}
The RG approach has the advantage that the nature of the low momentum
interaction can be revealed by studying the change of $\vlk$ with 
the cutoff scale. In this way, the RG evolution  
clearly provides evidence for the separation of scales in the 
two-nucleon problem, see also~\cite{Vlowk}. Furthermore, 
we will show that it also provides physical insight into the low 
momentum interaction. 

\begin{figure}[t]
\vspace{-8mm}
\begin{center}
\includegraphics[scale=1.15,clip=]{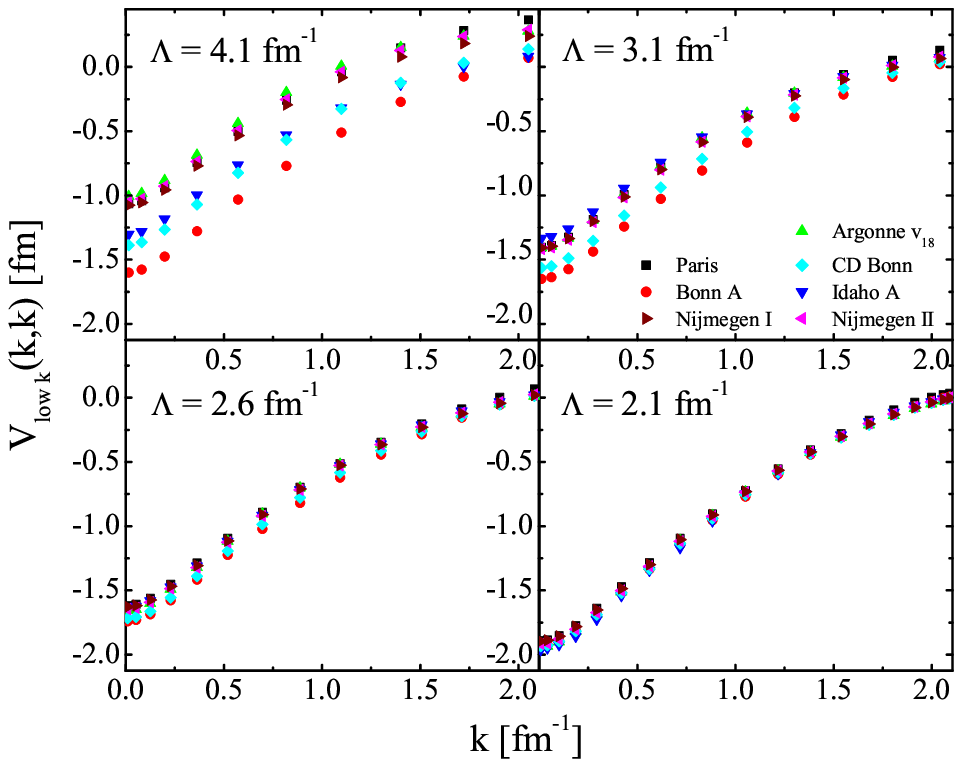}
\end{center}
\vspace{-8mm}
\caption{The collapse of the diagonal momentum-space matrix elements
of $\vlk$ as the cutoff is lowered to $\Lambda = 2.1 \, \text{fm}^{-1}$ 
in the $^3$S$_1$ partial wave.}
\label{collapse3s1}
\vspace{-8mm}
\begin{center}
\includegraphics[scale=1.15,clip=]{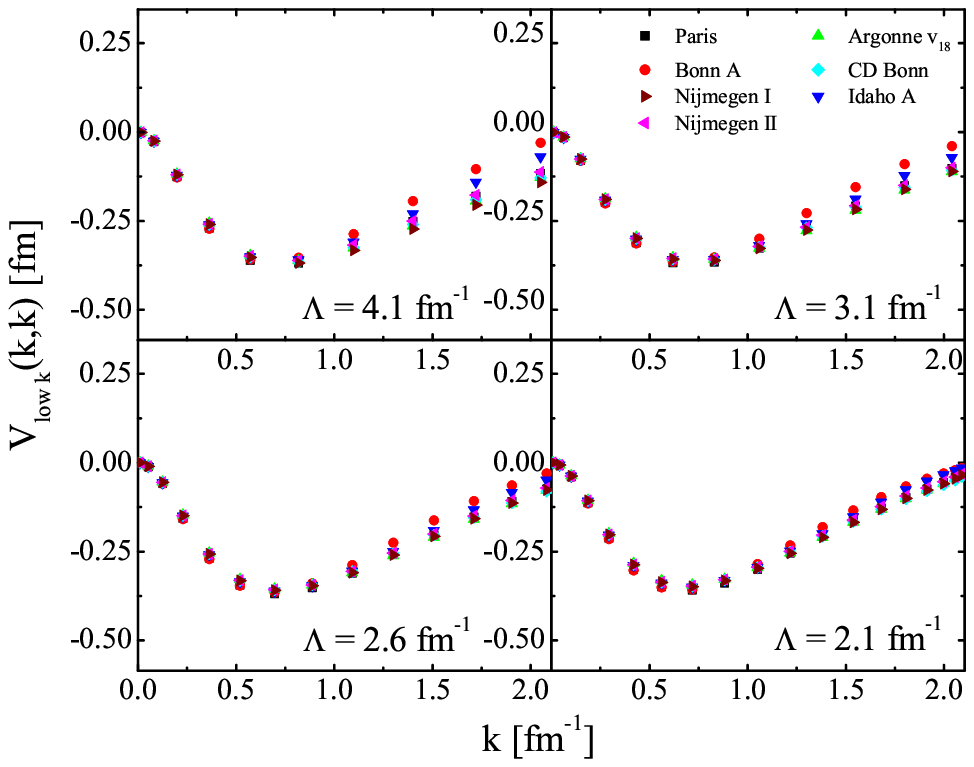}
\end{center}
\vspace{-8mm}
\caption{The collapse of the diagonal momentum-space matrix elements
of $\vlk$ as the cutoff is lowered to $\Lambda = 2.1 \, \text{fm}^{-1}$
in the coupled $^3$S$_1$-$^3$D$_1$ block.}
\label{collapse3sd1}
\end{figure}
In the evolution of the $^1$S$_0$ matrix elements shown in 
Fig.~\ref{collapse1s0}, we find that the model independence of $\vlk$
is reached at somewhat larger values of the cutoff around 
$\Lambda \sim 3.5 \, \text{fm}^{-1}$. This scale lies below the 
$\omega$ and $\rho$ meson masses, and consequently the details of 
the repulsive core are not resolved in the effective theory. Also note 
that for $\Lambda \gtrsim 3.0 \, \text{fm}^{-1}$, the Idaho potential 
is not renormalized, since it is an EFT potential without high momentum 
components by construction. 

Conversely, we find in Fig.~\ref{collapse3s1} 
that the collapse in the $^3$S$_1$ 
partial wave occurs at a comparatively lower value for the cutoff of
$\Lambda \sim 2.1 - 2.6 \, \text{fm}^{-1}$. This arises from the fact  
that the tensor force is operative in the triplet S-wave. Since
$\vlk$ acquires a large second-order renormalization from the tensor 
force, one indeed expects that $\vlk$ becomes model-independent at a 
lower cutoff in the $^3$S$_1$ partial wave. 

\begin{table}[t]
\begin{center}
\begin{tabular}{@{\hspace{0.5mm}}c@{\hspace{1mm}}c@{\hspace{1mm}}c@{\hspace{1mm}}c@{\hspace{1mm}}c@{\hspace{1mm}}c@{\hspace{1mm}}c@{\hspace{1mm}}c@{\hspace{0.5mm}}}
\hline
$E_{\text{D}} \: [\text{MeV}]$ & Paris & Bonn A & Nijm. I &
Nijm. II & Arg. $v_{18}$ & CD Bonn & Idaho A \\
\hline\hline
$V_{\text{NN}}$ & $-2.2218$ & $-2.2242$ & $-2.2246$ & $-2.2242$ & $-2.2247$ &
$-2.2238$ & $-2.2242$ \\
quoted & $-2.2249$ & $-2.22452$ & $-2.224575$ & $-2.224575$ & $-2.224575$ &
$-2.224575$ & $-2.2242$ \\
$\vlk$ & $-2.2218$ & $-2.2242$ & $-2.2246$ & $-2.2242$ & $-2.2247$ & $-2.2238$ & 
$-2.2242$ \\ 
\hline
\end{tabular}
\vspace*{0.5cm}
\end{center}
\caption{Comparison of the deuteron binding energy $E_{\text{D}}$ for
$\vlk$ and the bare interactions. We note that all $\vlk$ are derived 
with the same momentum mesh. For the values reported here, we have chosen
not to optimize the mesh for each bare interaction separately.}
\label{deuteronbe}
\end{table}
\begin{figure}[t]
\vspace{-8mm}
\begin{center}
\includegraphics[scale=1.15,clip=]{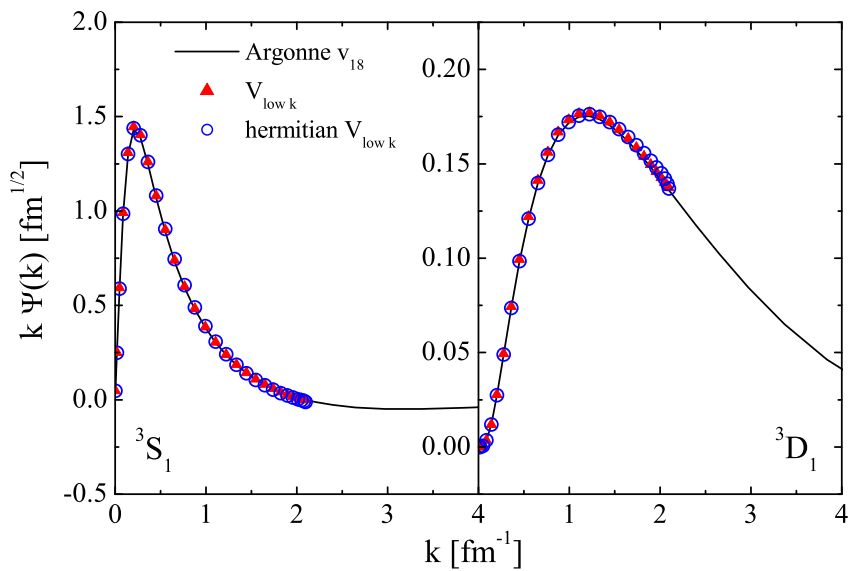}
\end{center}
\vspace{-8mm}
\caption{Comparison of the momentum-space S and D-state deuteron 
wave functions calculated from $\vlk$, the hermitian $\vlkb$ and the
bare $V_{\text{NN}}$. Results are shown for a cutoff of $\Lambda = 2.1
\, \text{fm}^{-1}$ and the Argonne $v_{18}$ potential.}
\label{deuteronwf}
\end{figure}
The nature of the tensor interaction of $\vlk$ can be studied by considering
the couping between the $^3$S$_1$ and $^3$D$_1$ partial wave. In this
channel, only the tensor force enters. The results for the RG
evolution in the $^3$S$_1$--$^3$D$_1$ block are presented in 
Fig.~\ref{collapse3sd1}. We observe that the realistic models are 
practically identical for momenta below $k \lesssim 0.7 \,
\text{fm}^{-1}$, which corresponds to the pion mass. This clearly
demonstrates that the tensor part of the input models differs
significantly for momenta above $m_\pi$. However, the effects of the
different high momentum tensor components on the low momentum physics
can be taken into account by employing a renormalized interaction, 
which in turn becomes model-independent at a cutoff of $\Lambda \sim
2.1 \, \text{fm}^{-1}$. 

The RG decimation to $\vlk$ preserves the deuteron binding energies of
the bare interactions. In Table~\ref{deuteronbe}, we give the binding
energies obtained from the $\vlk$, and verify that the deuteron poles
are in fact reproduced in the effective theory. We note that we have not
optimized the momentum mesh for each input $V_{\text{NN}}$, 
which would further improve the agreement with the quoted values.  
Referring to 
Fig.~\ref{deuteronwf}, the low momentum components of the bare 
deuteron S and D-state wave functions are RG invariant as well. 
Moreover, although the hermitian $\vlkb$ does not exactly preserve 
the low momentum projections of the wave functions, we find in 
Fig.~\ref{deuteronwf} that the deuteron wave functions calculated 
from $\vlk$ and the hermitian $\vlkb$ are practically identical. 

\begin{figure}[t]
\vspace{-8mm}
\begin{center}
\includegraphics[scale=1.15,clip=]{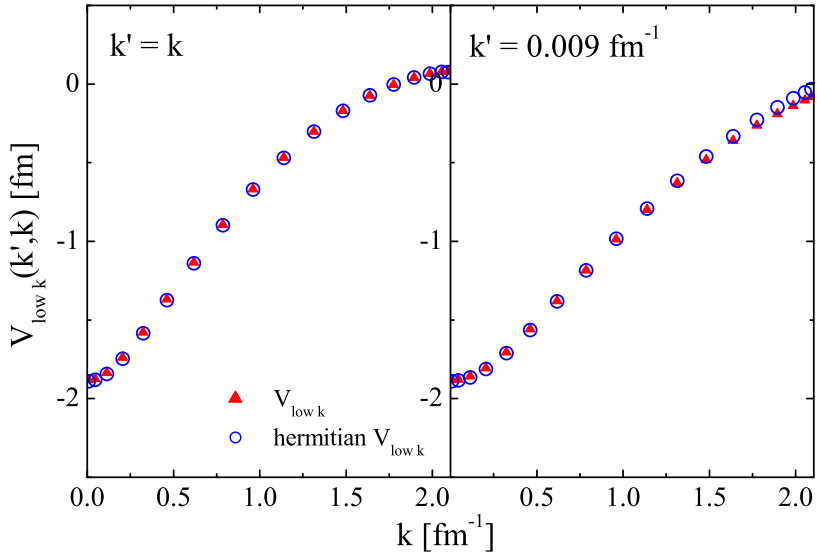}
\end{center}
\vspace{-8mm}
\caption{Comparison of the diagonal (left) and off-diagonal (right) matrix
elements of $\vlk$ and the hermitian $\vlkb$ in the $^1$S$_0$ partial wave.
Results are shown for a cutoff of $\Lambda = 2.1 \, \text{fm}^{-1}$ and 
the Argonne $v_{18}$ potential.}
\label{hermcomp}
\end{figure}
Finally, we note that the same general results hold for both $\vlk$
and $\vlkb$, and that the numerical differences are quite small. 
This is demonstrated as an example for the $^1$S$_0$ partial wave 
in Fig.~\ref{hermcomp}. It can also be seen that the differences are
as expected most noticeable for the far off-shell parts of the low
momentum interaction.

\subsection{Phase shift equivalence of the low momentum interaction}

The model-independent low momentum interaction reproduces the experimental 
phase shift data with similar accuracy as the high precision nucleon-nucleon
potentials. Therefore, $V_{\text{low k}}$ fits the measured phase shifts 
for laboratory energies below $E_{\text{lab}} = 350 \, \text{MeV}$ 
with a similar $\chi^2/\text{datum} \approx 1 - 2$, depending on which
conventional potential model $V_{\text{NN}}$ is used. 
Since the hermitian $\overline{V}_{\text{low k}}$ 
preserves this content as
well as the hermiticity of the Hamiltonian, we demonstrate the high
accuracy of the hermitian $\overline{V}_{\text{low k}}$ here. In 
Fig.~\ref{vlowkps1} and Fig.~\ref{vlowkps2}, the neutron-proton 
phase shifts and the mixing parameters in coupled channels for all 
partial waves $J \leqslant 4$ are given. For clarity of the Figures, we 
show these for the low momentum $\overline{V}_{\text{low k}}$ derived 
from the CD Bonn potential only. Since both the hermitian and non-hermitian 
low momentum interactions preserve the phase shifts of the input 
models, results of the same accuracy are derived from the various 
other models. For comparison, the phase shifts and mixing
parameters calculated from the bare CD Bonn potential and 
the results of the Nijmegen multi-energy phase shift 
analysis~\cite{PWA,NNonline} are included in Fig.~\ref{vlowkps1} and 
Fig.~\ref{vlowkps2}.

We conclude that the low-energy scattering data can be equally well
reproduced by an effective low momentum interaction $V_{\text{low k}}$
restricted to the Hilbert space of low momentum modes, $k <
\la_{\text{data}}$. This is achieved with a model-independent 
interaction, without introducing further assumptions on the details of the
short-distance dynamics.

\section{Summary and advantages}

We have shown that the RG decimation of different high precision
nucleon-nucleon interactions to low momenta leads to a unique 
interaction, called $V_{\text{low k}}$. This result is obtained by
truncating the full Hilbert space to an effective space, which
consists of momentum components up to the scale of the constraining 
low-energy data. The differences in the realistic interactions arise
from the assumed high momentum dynamics. By restricting the
interaction to the low momentum space, the details of the
short-distance physics are not resolved, and the detail-independent 
effects of the high momentum modes on the low momentum observables 
are included in the renormalized effective interaction. The existence 
of a nearly universal $V_{\text{low k}}$ demonstrates that 
it is possible to separate the physics of low momentum nucleons 
from the ambiguous short-distance parametrizations used in 
the realistic potential models. 

The RG invariance of the effective theory guarantees that the 
HOS $T$ matrix and the deuteron properties of the full theory 
are exactly preserved in the model space. 
Consequently, $V_{\text{low k}}$ is phase shift equivalent 
to the high precision potential models and reproduces the experimental 
elastic scattering data with similar accuracy. 
A further refinement transforms $\vlk$ into a hermitian low 
momentum interaction $\vlkb$, which remains model-independent and
preserves the low-energy on-shell amplitudes of the bare interactions.  
In this manner, the resulting RG decimation incorporates the 
model-independent effects of the short-distance physics in the 
low momentum interaction, while the model-dependent effects are 
filtered out.

We believe that the low momentum interaction $V_{\text{low k}}$ 
satisfies the first of the two requirements of a microscopic 
many-body theory given in the Introduction. Namely, that the input 
interaction is, as best as possible, based on the available low-energy data. 
A necessary condition for the collapse of the low momentum interactions
is the separation of long from short-distance scales in the nuclear 
problem. In addition, we have argued that the accuracy of the 
reproduction of the phase shifts drives the collapse of the 
on-shell components of the low momentum interactions. Furthermore, the
evolution of $\vlk$ with the cutoff provides physical insight into 
the separation of scales in the nuclear problem, the effects of the 
short-range repulsion in the renormalization, and the tensor content 
of the low momentum interaction.

$\vlk$ sums high momentum ladders in free space. Therefore, the low 
momentum interaction is considerably softer than the bare interactions 
and does not have a repulsive core. As a consequence, when used as the 
microscopic input in the many-body problem, $\vlk$ does not lead to
strong high momentum scattering effects in the particle-particle channel, 
which necessitates a Brueckner resummation or short-range
correlation methods for the conventional models.

The low momentum interaction should be regarded as a new input potential 
for many-body calculations. However, we emphasize that in contrast to 
the Brueckner $G$ matrix, $\vlk$ does not have high momentum components 
above the cutoff and therefore one does not have to be concerned about double
counting the high momentum contributions. In EFT, this is understood 
easily, since the ultraviolet divergences can be renormalized in free 
space and the many-body dynamics does not lead to new ultraviolet 
divergences. More precisely, in any many-body diagram, the excitations 
to high momentum states that are absorbed in $\vlk$ vanish due to 
the momentum space cutoff by construction. Using $\vlk$ is simply
analogous to using the relatively soft CD Bonn interaction, instead of 
the stronger Argonne $v_{18}$ potential.

Since $\vlk$ accounts for a large part of the phase space in the 
many-body system, it will be interesting to explore whether calculations 
starting with $\vlk$ could be organized in a more perturbative fashion.
Encouraging results for a perturbative expansion of the shell model 
effective Hamiltonian in few-body systems has been demonstrated for 
the deuteron in~\cite{Haxton2}. For larger systems, this idea more 
relies on a geometrical phase space argument.

\begin{ack}
We thank Gerry Brown, Bengt Friman and Dick Furnstahl for helpful 
discussions. This work was supported by the US-DOE Grant 
DE-FG02-88ER40388, the US-DOE Grant DE-FG03-00ER41132, the 
NSF under Grant No. PHY-0098645 and by an Ohio State University 
Postdoctoral Fellowship.
\end{ack}

\newpage

\appendix
\section{Diagonal matrix elements of the hermitian $\vlkb$}

\begin{figure}[b]
\vspace{-8mm}
\begin{center}
\includegraphics[scale=1.14,clip=]{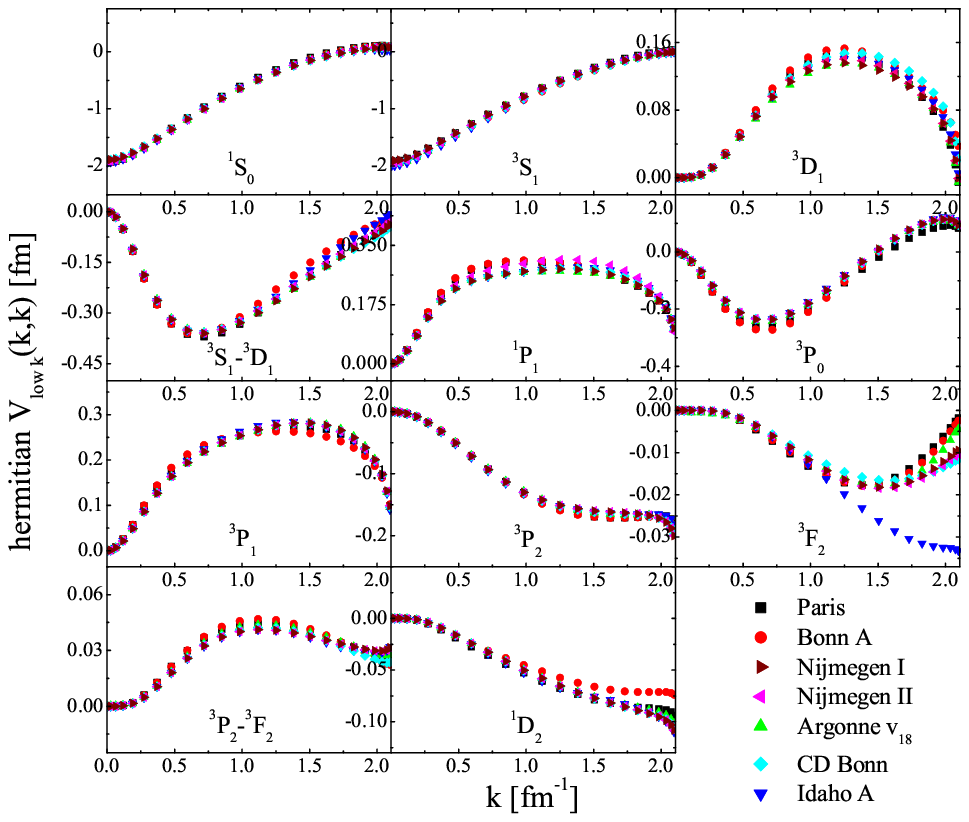}
\end{center}
\vspace{-8mm}
\caption{Diagonal momentum-space matrix elements of the hermitian 
$\vlkb$ obtained from the different potential models for a cutoff 
$\Lambda = 2.1 \, \text{fm}^{-1}$. Results are shown for the 
partial waves $J \leqslant 4$.} 
\label{hermdiag1}
\vspace{-6mm}
\begin{center}
\includegraphics[scale=1.14,clip=]{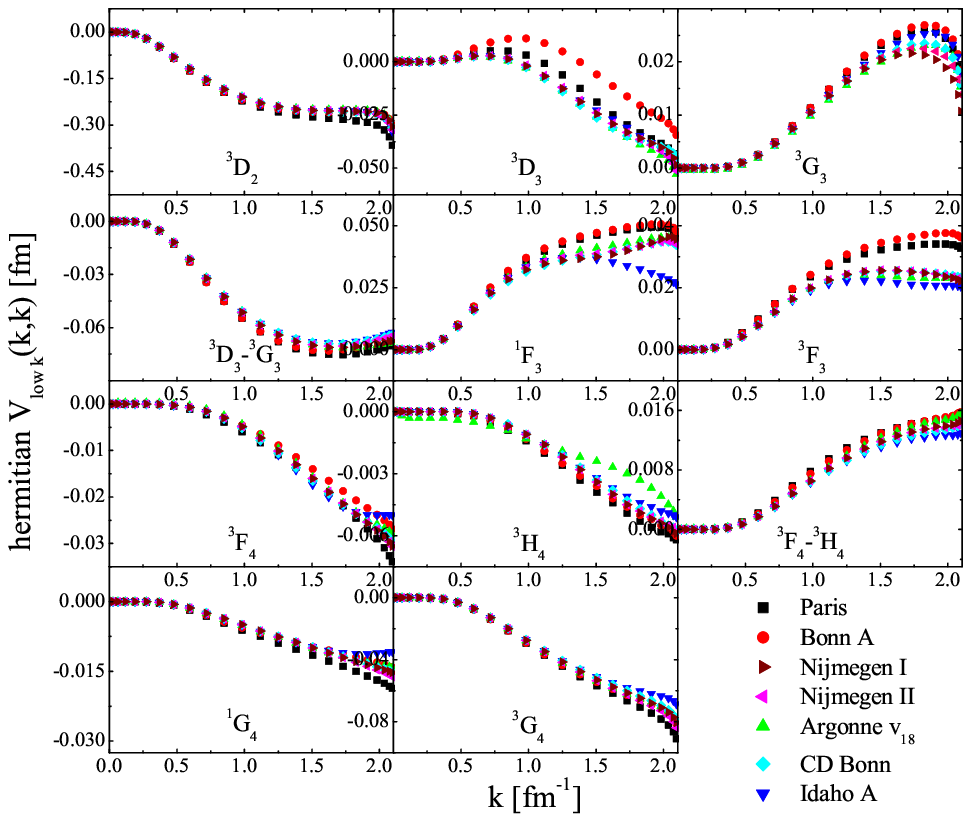}
\end{center}
\vspace{-8mm}
\caption{The diagonal elements of the hermitian $\vlkb$ are
continued from Fig.~\ref{hermdiag1}.}
\label{hermdiag2}
\end{figure}
\end{document}